\documentclass[useAMS,usegraphicx,usenatbib]{mn2e} 
\usepackage{ulem}
\usepackage{color}
\usepackage{graphics,graphicx}
\usepackage{times} 
\usepackage{amssymb}
\usepackage{amsmath}
\usepackage{lscape}
\usepackage{url}
\newif\ifAMStwofonts
\AMStwofontstrue
\def\rg{${\it r}_{\rm g}$}
\def\rin{${\it r}_{\rm in}$}
\def\laor{\rm{\sc LAOR}}
\def\phabs{\rm{\sc PHABS}}
\def\diskbb{\rm{\sc DISKBB}}
\def\ezdiskbb{\rm{\sc EZDISKBB}}
\def\cflux{\rm{\sc CFLUX}}
\def\steppar{\rm{\sc STEPPAR}}

\def\diskpn{\rm{\sc DISKPN}}

\def\reflionx{\rm{\sc REFLIONX}} 

\def\kdblur{\rm{\sc KDBLUR}} 

\def\nh{${\it N}_{\rm H}$} 
\def\ka{{\rm K}$\alpha$}

\def\epicpn{{\it EPIC}{\rm-pn}}
\def\epicmos1{{\it EPIC}{\rm-MOS1}}
\def\epicmos2{{\it EPIC}{\rm-MOS2}}
\def\epicmos{{\it EPIC}{\rm-MOS}}

\def\ixo{{\it IXO}}
\def\chandra{{\it Chandra}}
\def\xmm{{\it XMM-Newton}}
\def\swift{{\it SWIFT}}
\def\asca{{\it ASCA}}
\def\integral{{\it INTEGRAL}}
\def\suzaku{{\it Suzaku}}
\def\bepposax{{\it BeppoSAX}}
\def\rxte{{\it RXTE}}
\def\rxtepca{{\it RXTE/PCA}}

\def\xspec{\hbox{\sc XSPEC}}
\def\xspecv{{\sc XSPEC}{\rm\thinspace v\thinspace 12.5.0}}

\def\heasoftv{\hbox{\rm HEASOFT\thinspace v6.6.2}}
\def\xselect{\hbox{\rm XSELECT}}
\def\ftool{\hbox{\rm FTOOL}}
\def\grppha{\hbox{\rm GRPPHA}}
\def\addascaspec{\hbox{\rm ADDASCASPEC}}

\def\ftest{\hbox{\rm FTEST}}

\def\s{\hbox{$\rm\thinspace s$}}
\def\ks{\hbox{$\rm\thinspace ks$}}

\def\deg{$^{\circ}$}  

\def\cm{\hbox{$\rm\thinspace cm$}}

\def\kpc{\hbox{$\rm\thinspace kpc$}}

\def\pcmsq{\hbox{$\rm\thinspace cm^{-2}$}}

\def\ev{\hbox{$\rm\thinspace eV$}}
\def\kev{\hbox{$\rm\thinspace keV$}}

\def\ergpcmsqps{\hbox{$\rm\thinspace erg~cm^{-2}~s^{-1}$}}

\def\msun{\hbox{$\rm\thinspace M_{\odot}$}}

\def\gx{\hbox{\rm GX 339-4}}
\def\j1655{\hbox{\rm J1655-40}}
\def\j1753{\hbox{\rm J1753.5-0127}}
\def\j1118{\hbox{\rm XTE J1118+480}}

\def\j1650{\hbox{\rm J1650-500}}
\def\j1749{\hbox{\rm J17497-2821}}

\begin{document}
\title[Accretion disks in the low-hard state] {Black hole accretion disks in the canonical low-hard state} \author[Reis, Miller, \& Fabian] {\parbox[]{6.in}
  {R.~C.~Reis $^{1}$\thanks{E-mail:
      rcr36@ast.cam.ac.uk}, A.~C.~Fabian$^{1}$, and J.~M.~Miller$^{2}$\\ } \\
  \footnotesize
  $^{1}$Institute of Astronomy, Madingley Road, Cambridge, CB3 0HA\\
  $^{2}$Department of Astronomy, University of Michigan, Ann Arbor}
\maketitle

\begin{abstract} 

Stellar-mass black holes in the low-hard state may hold clues to jet
formation and basic accretion disk physics, but the nature of the
accretion flow remains uncertain.  A standard thin disk can extend
close to the innermost stable circular orbit, but the inner disk
may evaporate when the mass accretion rate is reduced. Blackbody-like continuum emission and dynamically-broadened iron emission lines provide independent means of
probing the radial extent of the inner disk.  Here, we present an X-ray study
of eight black holes in the low-hard state.  A thermal disk continuum with a colour
temperature consistent with $L \propto T^{4}$ is clearly detected in
all eight sources, down to $\approx$5$\times10^{-4}L_{Edd}$.  In six
sources, disk models exclude a truncation radius larger than 10\rg.
Iron-\ka\ fluorescence line emission is observed in half of the
sample, down to luminosities of $\approx$1.5$\times10^{-3}L_{Edd}$.
Detailed fits to the line profiles exclude a truncated disk in each
case.  If strong evidence of truncation is defined as (1) a non-detection of a broad iron line, {\it and} (2) an inner disk temperature much cooler than expected from the ${\rm L}
\propto {\rm T}^{4}$ relation, none of the spectra in this sample
offer strong evidence of disk truncation.  This suggests that the
inner disk may evaporate at or below $\approx$1.5$\times10^{-3}L_{Edd}$.

\end{abstract}

\begin{keywords} X-rays: -- accretion disk    

\end{keywords}

\section{Introduction}

The innermost regions of accreting black holes and the underlying
physics of the accretion process can be successfully studied using the
X-ray spectra of black hole binaries. Over the past years, such
studies have revealed that Galactic black hole binaries (GBHBs)
radiate in various distinct spectral states characterised by the
relative strength of their soft and hard X-ray emission (McClintock \&
Remillard 2006).

The X-ray spectrum of black hole binary systems in their low-hard
state is dominated by a powerlaw with relatively low luminosity
($\le$0.05$L_{Edd}$; Maccarone 2003), a photon index $\Gamma$ in the
range $\approx$1.4--2 and an exponential cut-off at about 100\kev. This
is in contrast with the high-soft state, where the spectrum is
dominated by a quasi-black body component with a characteristic
temperature of $\sim$1\kev, and the highly luminous very--high
state ($L$$\approx$$L_{Edd}$) with a powerlaw component having a flux
comparable to that of the soft blackbody. In the latter cases, the
powerlaw does not show evidence for a high--energy
roll-over. Superimposed on the continuum is the presence of various
reflection features (Ross \& Fabian 1993; Miller 2007) with the
dominant component being the Fe-\ka\ emission line. In the inner
regions of the accretion flow the reflection spectrum appears blurred
due to the combination of various relativistic effects. The degree of
gravitational blurring is strongly dependent on the inner radius of
the reflecting material (Fabian et al. 1989; Laor 1991).

Theoretically, the soft, quasi-black body component observed in the
high-soft and very-high state is generally agreed to originate in a
standard, optically thick, geometrically thin accretion disk (Shakura
\& Sunyaev 1973) extending to the innermost stable circular orbit
(ISCO). For a non-rotating, Schwarzschild, black hole this radius is
equal to 6\rg\ where
\rg{\thinspace = $GM/c^2$}, and decreases to $\approx$1.2\rg\ for a
maximally-rotating, Kerr black hole (Thorne 1974). The powerlaw
component dominating the X-ray spectra in the low-hard state, on the
other hand, is believed to be produced by the inverse Compton
scattering of soft photons in a thermal, optically thin region
(Shapiro et al. 1976; Sunyaev \& Titarchuk 1980).  However, there is
no general consensus on the geometry of this optical--thin region.  In
the accretion disk-corona model the hard powerlaw comes from either a
patchy corona, possibly powered by magnetic flares (Di Matteo, Celotti
\& Fabian 1999; Beloborodov 1999; Merloni, Di Matteo \& Fabian 2000;
Merloni \& Fabian 2001) or the base of a centrally located jet
(Merloni \& Fabian 2002; Markoff \& Nowak 2004; Markoff, Nowak \&
Wilms 2005).  In both cases the thin accretion disk extends
close to the ISCO. An alternative model has the thin disk truncated at
large distances from the black hole (Narayan \& Yi 1995; Esin et
al. 1997) with the central region replaced by an advection-dominated
accretion flow (ADAF).

In both interpretations for the geometry of the accretion disk in the
low-hard state it is expected that the thermal-disk component will
have a low effective temperature. The low mass accretion rates, and
corresponding low luminosity, observed in systems in the low-hard
state implies a peak disk temperature $<0.4$\kev. In the case of a
geometrically thin disk extending close to the ISCO, this temperature
closely follows the $L\propto T^4$ relation. The expected temperature
departs dramatically from this relation when the disk is truncated at
the distances predicted by the ADAF models (see e.g. McClintock et
al. 2001). A thermal component has been observed in a number of
sources in the low-hard state. In particular, \xmm/\rxte\ observations
of \gx\ (Miller et al. 2006; Reis et al. 2008) and Swift J1753.5-0.127
(Miller, Homan \& Miniutti 2006; Reis et al. 2009b) as well as
\chandra\ observation of XTE J1118+480 (Reis, Miller \& Fabian 2009)
have shown that the accretion disk in these sources are consistent
with extending close to the ISCO. Evidence for the disappearance of
this component has usually been the result of observations made with
\rxte\ which lacks the low energy coverage.

As mentioned above, a further feature in the X-ray spectra of BHBs are
the various reflection signatures associated with the reprocessing of
hard radiation by the cool accretion disk. The most prominent of these
features is often the broad, skewed Fe-\ka\ line observed in a number of
BHBs (Miller 2007, 2009), Seyfert galaxies (Tanaka et al. 1995;
Fabian et al. 2009) and accreting neutron stars (Cackett et al. 2008;
Reis, Fabian \& Young 2009).  New fits to some ultra-luminous X-ray sources (ULXs;
Caballero-Garcia \& Fabian 2009) spectra suggest that a disk
reflection may also be observed, though more sensitive spectra are
needed to confirm this possibility. The strength of the reflection
features can be typified by the equivalent width of the iron-\ka\
line, $W_{K\alpha}$ which is expected to correlate linearly with the
reflection fraction $R = \Omega/2\pi$, where $\Omega$ is defined as
the solid angle covered by the accretion disk as viewed from the hard
X-ray source. George \& Fabian (1991) have shown that for a neutral
accretion disk with solar abundances the reflection fraction closely
follows $R\approx W_{K\alpha}/180\ev$. For stellar mass black holes in
the low-hard state, $R$ is observed to be below $\sim 1$. This
``weak'' reflection can be interpreted as either a recession in the
accretion disk, as in the ADAF interpretation (Esin et al. 1997), a
highly ionised inner disk (Ross, Fabian \& Young 1999) or mildly
relativistic motion of the hot corona away from the disk (Beloborodov
1999). Using the latter model, Beloborodov showed that reflection
fractions as low as $\approx 0.3$ can be be obtained in the low-hard
state without invoking disk truncation.

In the following chapters we will present a systematic analysis of a
sample of eight stellar mass black hole binaries in the low-hard state
with the goal of better understanding the accretion disk and flow
geometry of such systems. We find that for the sources observed in
this work the accretion disks are consistent with extending close to
the ISCO and suggest a set of observational criteria for the support
of disk truncation. We start in \S{\thinspace\ref{observation}} with
an introduction to the various systems investigated. This is then
followed by a detailed analyses of the property of the thermal
component (\S\S{\thinspace\ref{analyses_mcd}}--\ref{innerrad_diskpn})
and the reflection component (\S\ref{innerrad_reflection}). Our
results are summarised in \S{\thinspace\ref{discussion}}.

\section{Observations and Data reduction}
\label{observation}
Our aim is to present a selection of high quality spectra of various
X-ray binaries in the low-hard state. In such a state the temperature
of the accretion disk is usually observed to be below $\approx$0.4\kev\
and for this reason a strong requirement of our selection is that the
data extends to a similar energy range. Observations made {\it solely}
with the {\it Rossi X-ray Timing Explores} (RXTE) are thus not used in
this work due to its low energy cut-off of approximately 3\kev. The
bulk of the work presented here utilises observations made with either
\xmm\ or \suzaku, with two further sources observed with \chandra\ and
\swift.  In what follows we introduce the various sources which meet
this criterion and describe the observations and data reduction in
detail.

\subsection{GX{\thinspace 339-4}}

GX{\thinspace 339-4} is a dynamically constrained black hole binary
located at a distance of $8\pm2$\kpc\ (Zdziarski et
al. 2004). Although its mass is not yet known, it is likely that \gx\
is amongst the more massive of the stellar mass black hole
sources. The mass function of the system has been constrained to
$\approx$6$\msun$ (Hynes et al. 2003 and Munoz-Darias et al. 2008) and
results derived from radio observations suggests a low ($\theta \le
30$\deg) inclination for the inner disk (Gallo et al. 2004). Similar
low inclinations have also been found when modelling the reflection
features present in the spectrum of \gx\ in outburst (Miller et
al. 2004). The apparent low inner disk inclination mentioned above
does not necessarily imply a massive ($>30\msun$) stellar mass black
hole since it is possible that the inner disk is warped similarly to
that observed in GRO{\thinspace J1655-40} (Martin, Tout \& Pringle
2008) and V4641{\thinspace Sgr} (Martin, Reis \& Pringle 2008). For
the purpose of this work we assume a black hole mass in the range of
10--20$\msun$ and an inner disk inclination between 10--30 degrees. We
note here that throughout this paper we will adopt the largest range
for the physical parameters (mass, distance and inclination) available
in the literature for the various sources, unless the parameters have
been well constrained.

Prior analyses of \gx\ in both the very-high (Miller et al. 2004) and
intermediate (Miller et al. 2008) state have shown that the accretion
disk extends close to 2\rg\ in those states. A similar conclusion has
been reached by jointly fitting the reflection features present in the
\xmm\ spectra of both the low-hard and the very-high state (Reis et
al. 2008). In this work, the low-hard state spectrum of \gx\ described
in Miller et al. (2004) is re-analysed using the more stringent
reduction procedure detailed in Reis et al. (2008). We restrict our
spectral analyses of the \xmm\ \epicmos\ data to the 0.5--10.0\kev\
energy range. The simultaneous \rxte\ PCA spectrum is fit in the
standard 3.0--25.0\kev\ energy range with 0.6 per cent systematic
error and an edge at 4.78\kev\ ($\tau = 0.1$) to account for a strong
Xe{\thinspace L} absorption feature. The HXTE spectrum is fit in the
25.0--100\kev\ range.  When fitting the RXTE spectra along with that
of \xmm, a joint fit is achieved by allowing a normalisation constant
to float between the various spectra. All parameters were tied between
observations unless stated otherwise. \xspecv\ (Arnaud 1996) was used
to analyse all spectra presented in this paper.  The quoted errors
correspond to a 90 per cent confidence level for one parameter of
interest unless stated otherwise.

\subsection{XTE{\thinspace J1650-500}}

Optical observations have constrained the mass function of
XTE{\thinspace J1650-500} (hereafter J1650) to $2.73\pm0.56\msun$
(Orosz et al. 2004). Together with a lower limit on the inclination of
50\deg$\pm3$\deg\ this sets an upper limit to the mass of the central
source of approximately 7.3$\msun$. A recent analysis made by
Shaposhnikov \& Titarchuk (2009) has placed a strong constraint on
the mass of the central black hole of $9.7\pm1.6\msun$ based on the
scaling properties of quasi-periodic oscillations (QPOs). The
discrepancy between their results and that of Orosz et al. (2004) are
attributed to a possible mis-classification of the spectral class of
the optical companion. Due to the uncertainty in the mass of J1650 we
conservatively assume that it lies between 5.3 and 11.3$\msun$. The
distance to J1650 has recently been estimated at $2.6\pm0.7$\kpc\
(Homan et al. 2006).

\xmm\ observed J1650 during the transition from the low-hard to the
very-high state in 2001 (Miller et al. 2002).  The source was
mistakenly identified as being in the very-high state due to the
presence of an unusually strong (at the time) iron line
emission. However, the cold ($\approx$0.3\kev) thermal component and low
luminosity likely place the observation in a rising phase of the
low-hard state (see e.g. Rossi et al. 2005).  The presence of a broad,
asymmetric Fe emission line led Miller et al. to suggest that the black
hole in J1650 is rapidly rotating. This was later confirmed by
Miniutti, Fabian \& Miller (2004) using \bepposax\ observations taken
both before and after that of \xmm. Contrary to this interpretation,
Done \& Gierlinski (2006) showed that the bright, low-hard state,
\bepposax\ observation is also consistent with a disk truncated at
$\sim$10--20\rg\ if there is resonance iron-K absorption from an
outflow disk wind with a velocity of $\approx$0.15c.

Here, we analyse the same \xmm\ \epicpn\ observation described in
Miller et al. (2002). The source was observed in the ``burst'' mode
with the ``thin'' optical filter in place. The spectrum was extracted
using a box region centred on the source with a 20 pixel
width. Background events were extracted from a 10 pixel wide box
adjacent to the source. Single and double events were included in the
analysis. Response files were created in the standard way using the
tools rmfgen and arfgen. The \ftool\ \grppha was used to require at
least 20 counts per bin in all \xmm\ observations presented in this
paper. We restrict our study of the \xmm\ \epicpn\ data to the
0.6--10.0\kev\ energy range.

\subsection{Cygnus{\thinspace X-1}}

Cygnus{\thinspace X-1} (hereafter Cyg{\thinspace X-1}) is amongst the
better known and studied stellar mass black holes. Having been the
first object generally recognised as a black hole (Tananbaum et
al. 1972) it has received considerable attention. The distance to the
source was very early estimated at 2\kpc\ (Murdin \& Webster
1971). This has been confirmed more recently by Massey et al. (1995)
where a distance of $2.1\pm0.1$\kpc\ is found. The orbital inclination
of the system is thought to be in the range of 25--50 degrees
(Gierlinski et al. 1999). The uncertainty in this value follows the
fact that the accretion disk in Cygnus{\thinspace X-1} is thought to
be precessing (Stirling et al. 2001; Romero et al. 2002) with an
average angle to the line of sight of $\sim$30\deg\ (Fender 2001). The
mass of the black hole is also highly uncertain, partially due to the
uncertainty in the inclination. Although it is generally assumed to be
close to 10\msun, its value has been found by various authors to range
from 7--25$\msun$ (e.g. Ziolkowski 2005; Shaposhnikov \& Titarchuk
2009).

 Young et al. (2001) showed that the various spectra of
Cygnus{\thinspace X-1} in both the low-hard and the high-soft state
are consistent with the relativistic blurring of a thin, ionised
accretion disk extending close to the ISCO. This interpretation was
later challenged by Barrio, Done \& Nayakshin (2003). Using the high
energy coverage of \rxte\ the authors were unsuccessful in measuring
the high energy break assumed to be characteristic of the ionised
accretion disk model{\footnote {If the accretion disk is ionised
    solely due to the hard X-rays from magnetic flares then a high
    energy break is expected at the thermal temperature of the
    flares. However it is possible that the source of hard radiation
    contains a hybrid of thermal/non thermal electrons at which point
    a break is no longer necessary in the hard spectra.}}, and thus
argued for a truncated disk interpretation of the spectra.

Makishima et al. (2008) observed the low-hard state of
Cygnus{\thinspace X-1} in 2005 October with \suzaku. Although the data
suffered heavily from both pile up and telemetry saturation, the
authors found the presence of a strong soft component with a
temperature of $\approx$0.19\kev\ which was associated with an accretion
disk extending to within $\approx$250{\thinspace km}.  Assuming a mass of
10$\msun$ this is approximately 16\rg. We report on two recent
\suzaku\ observations where the effect of pile-up is less prominent.

Cygnus{\thinspace X-1} was observed with \suzaku\ on two occasions on
2009 April 8 (Obs ID 404075020; hereafter Obs 1) and April 14 (Obs ID
404075030; hereafter Obs 2) for approximately 50 and 33\ks\
respectively. The three operating detectors constituting the X-ray
Imaging Spectrometer (XIS; Koyama et al. 2007) were operated in the
``burst'' clock mode in both observations and employed the ``1/4
window'' mode. The front illuminated detectors (FI; XIS0 and XIS3)
were operated in the 2x2 and 3x3 editing mode, whereas the back
illuminated detector (BI; XIS1) was operated in the 3x3 editing mode
only. A total FI exposure of $\approx$10.7 and $\approx$8.6\ks\ for Obs 1
and 2 respectively were available for each XIS camera. The
corresponding BI exposures were 5.3 and 4.4\ks. Using the latest
\heasoftv\ software package we reprocessed the data from the Version 2
processing following the SUZAKU Data Reduction
Guide{\footnote{http://heasarc.gsfc.nasa.gov/docs/suzaku/analysis/}}. We
started by creating new cleaned event files using the tool {\rm
  ``xispi''} and the script {\rm ``xisrepro''} as well as the
associated screening criteria files.  \xselect\ was then used to
extract spectral products. In both observations, source events were
extracted from a region composed of the intersection of a rectangle of
size 150x1024 pixels with an annulus having 30 and 200 pixels inner
and outer radius respectively. Background events were extracted from a
circular region with a radius of 90 pixels away from the source.  We
used the script ``xisresp''{\footnote
  {http://suzaku.gsfc.nasa.gov/docs/suzaku/analysis/xisresp}} with the
``medium'' input to obtain individual ancillary response files (arfs)
and redistribution matrix files (rmfs). ``xisresp'' calls the tools
``xisrmfgen'' and ``xissimarfgen''. Finally, we combined the spectra
and response files from the two front-illuminated instruments (XIS0
and XIS3) using the \ftool\ \addascaspec.  This procedure was repeated
for each observation resulting in a total of four XIS spectra. The
\ftool\ \grppha\ was used to give at least 500 counts per spectral bin
in every \suzaku\ observation mentioned in this paper. The Hard X-ray
Detector (HXD) observations of Cyg{\thinspace X-1} will not be used in this work.
The FI-XIS spectra were fit in the 0.7--10.0\kev\ energy range. The
results obtained with the BI spectra are not presented in this work as
it is not expected to be different to that of the FI. In all fits to
\suzaku\ data presented in this paper the 1.9--2.5\kev\ energy range
was ignored due to the possible presence of unmodelled instrumental
features.

\subsection{SWIFT{\thinspace J1753.5-0127}}

Using multiwavelength observations, Cadolle-Bel et al. (2007) estimated
the distance to SWIFT{\thinspace J1753.5-0127} (hereafter J1753) to be
in the range of 4--8\kpc. This was based on the derived hydrogen
column density along the line of sight to the source and its known
latitude. From estimated bolometric flux and the assumption that the
source is radiating at less than 5 percent the Eddington luminosity,
the authors constrained the mass of J1753 to approximately
4--16$\msun$. This range is in agreement with the value of 12$\msun$
estimated by Zurita et al. (2008) based on the orbital period of
J1753.  In the same work Zurita et al. estimated a lower limit for the
distance of 7.2\kpc. For the purpose of our work we will assume a
conservative range of 4--16$\msun$ and 7.2--10\kpc\ for the mass and
distance respectively. The inner disk inclination was found from the
reflection features in the \xmm\ X-ray spectrum to be $55^{+2}_{-7}$
degrees (Reis et al. 2009b).

Using RXTE and XMM-Newton data, Miller, Homan \& Miniutti (2006)
showed the presence of a cool (kT $\approx$0.2\kev) accretion disk
extending close to the ISCO in the low-hard state of J1753.  The
presence of this cool accretion disk was later confirmed with \rxte\
observations during its 2005 outburst (Ramadevi \& Seetha 2007). Reis
et al. (2009b) have recently re-analysed the \xmm\ and \rxte\ spectra
of J1753 in the low-hard state and found that the disk likely extends
to \rin$=3.1^{+0.7}_{-0.6}$\rg. This strong constraint was based on
the self-consistent modelling of both the reflection features and the
soft thermal-disk component. In this work we further explore the
spectra described in Reis et al. (2009b) in the 0.5--10.0\kev\ energy
range.

\subsection{GRO{\thinspace J1655-40}}

GRO{\thinspace J1655-40} (hereafter J1655) has a compact object with a
mass constrained to $6.3\pm0.5\msun$ and orbital inclination of
70.2\deg$\pm$1.9\deg\ (Greene, Bailyn \& Orosz 2001). Using radio
observations, Hjellming \& Rupen (1995) found a distance of
$3.2\pm0.2$\kpc\ and an inner disk inclination of 85\deg$\pm$2\deg. In
this work we use an inclination of 70.2\deg\ with a lower and upper
limit of 68.3 and 87\deg\ respectively.

Using archival ASCA data of J1655 in outburst, Miller et al. (2005)
showed evidence of a highly skewed, relativistic line, and suggested
an inner accretion disk radius of $\approx$1.4\rg, indicative of a
rapidly spinning black hole. This was later confirmed by Diaz-Trigo et
al. (2007) using simultaneous \xmm\ and INTEGRAL observations of
J1655 during the 2005 outburst. Brocksopp et al. (2006) followed
the spectral evolution during the 2005 outburst using the \swift\
X-ray Telescope (XRT). They note the presence of a strong iron line at
$\approx$6.4\kev\ in the single observation of the low-hard
state{\footnote {Brocksopp et al. (2006) analysed twenty \swift\
    pointing covering the evolution from the low-hard, high-soft and
    very-high state, however only one pointing captured the source in
    the low-hard state.}}, however they find no evidence of a thermal
component. It is interesting that the neutral hydrogen column density
(\nh) remains at an average of $\approx$6.9$\times10^{21}\pcmsq$ for all
observations reported by Brocksopp et al. other than the single
low-hard state observation where the value of \nh\ drops to
$\approx$5.9$\times10^{21}\pcmsq$ (see their Tables 4 and 5). It is
plausible that a lower value of \nh\ can act to mask the presence of a
weak soft component when the signal to noise level is low (See
\S\thinspace\ref{analysis}).

Takahashi et al. (2008) have recently reported on the \suzaku\
observation of J1655 in the low-hard state where they find strong
evidence for both a broad iron line and a soft disk component with a
temperature of $\approx$0.2\kev. \suzaku\ observed J1655 in 2005
September 22, 07:32 UT for approximately 35\ks\ (Obs ID
100029010). The XISs were operated in the 3x3 and 2x2 editing mode
with the ``1/8 window'' employed. XIS0 was not operational during the
observation. The data were reduced in the standard way as described
above for Cyg{\thinspace X-1}. Source events were extracted from a circular
region 70 pixels in radius centred on the source. The background is
taken from a source-free region 50 pixels in radius towards the edge
of the chip. The HXD was operated in its normal mode. The appropriate
response and background files for XIS-nominal pointing were downloaded{\footnote{http://www.astro.isas.ac.jp/suzaku/analysis/hxd/}}
and the HXD/PIN data were reprocessed in accordance with the SUZAKU
Data Reduction Guide. The XIS and PIN data were fit in the 0.7--9.0
and 12.0--70.0\kev\ energy range respectively. A normalisation
constant of 1.16 was applied to the PIN data in accordance with the
latest \suzaku\ calibration report.

\subsection{XTE{\thinspace J1118+480}}

Dynamical measurements of XTE J1118+480 (hereafter J1118) have set a
strong constraint on the mass-function of $6.1\pm 0.3\msun$ (Wagner et
al. 2001; McClintock et al. 2001). Frontera et al. (2001) suggests
that the black hole in J1118 has a mass of 7--10$\msun$ with an
orbital inclination in the range of 60--80 degrees. A recent study by
Gelino et al. (2006) places a more stringent constraint of
$8.53\pm0.60\msun$ and 68\deg$\pm$2\deg\ for the mass and inclination
respectively. The authors find a distance of $1.72\pm0.1$\kpc\ to the
system. To our knowledge the broadest constraint on the distance to
J1118 was that imposed by McClintock et al. (2001) of
$1.8\pm0.6$\kpc. Since this range encompass all the latest predictions
we employ this value in the work that follows. The inclination and
mass is hereafter assumed to lie in the range 60--83 degrees{\footnote
  {The high value reflects the upper limit predicted by Wagner et
    al. (2001).}} and 7--10$\msun$ respectively.

  J1118 was observed in its LHS by \chandra, \rxte, UKIRT, EUVE, and the HST in 2000 as
part of a multiwavelength, multi-epoch observing campaign. Based
on these observation, McClintock et al. (2001) reported an
  apparent cool thermal component at $\approx$24\ev\ which was
  interpreted as originating in an accretion disk truncated at a
  radius greater than 70\rg. Reis, Miller \& Fabian (2009) have
  recently re-analysed the \chandra\ and \rxte\ data and shown that the accretion disk is
  consistent with extending close to the ISCO, contrary to previous
  results. The \chandra\ and \rxte\ spectrum of J1118 in the low-hard state, as
  described in Reis, Miller \& Fabian (2009) is used in the analyses
  that follows.

\subsection{IGR{\thinspace J17497-2821}}

\begin{figure*}
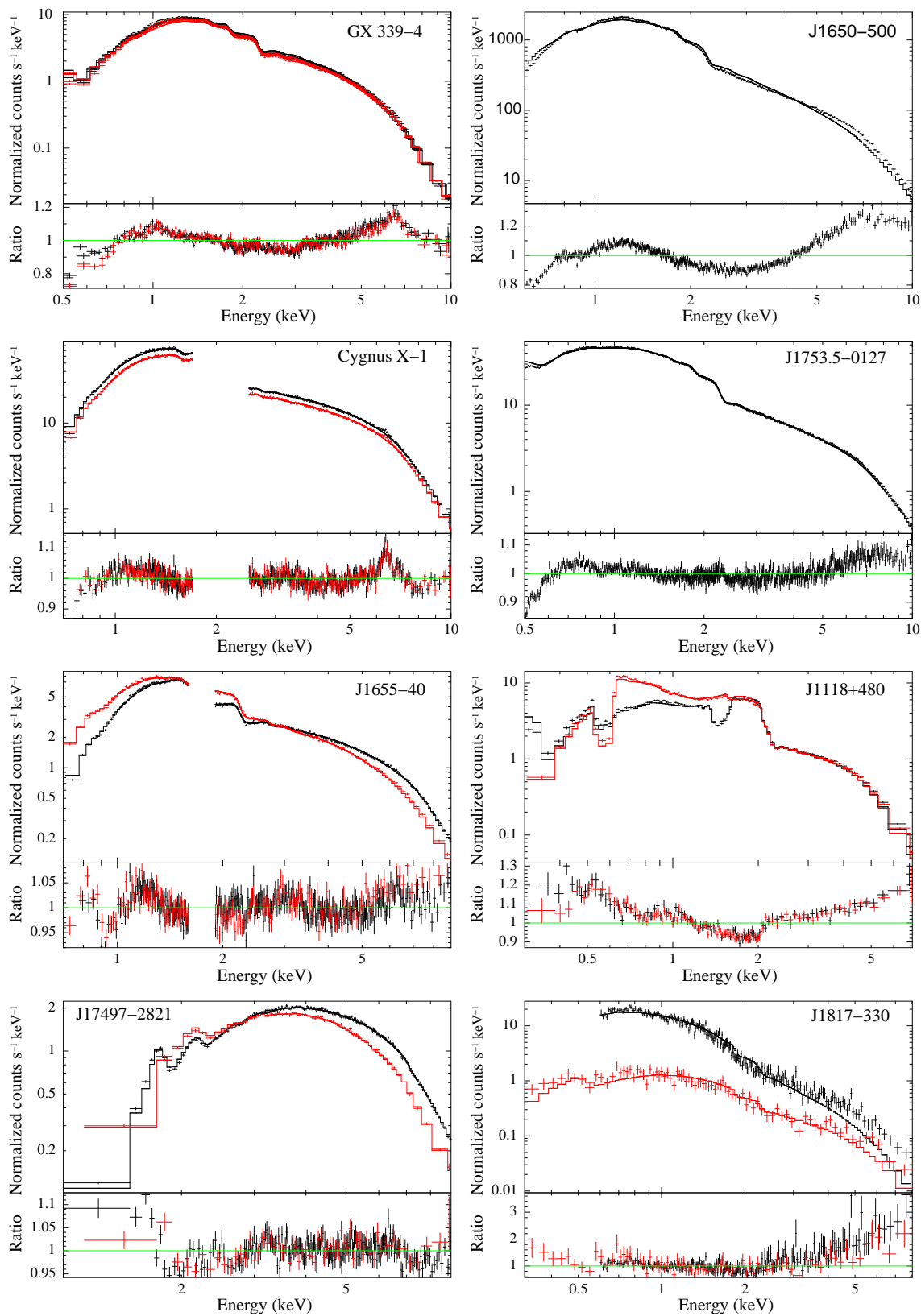

\begin{center}
\rotatebox{270}{
{\includegraphics[width=150pt]{figure_gx339_ratio_topl.ps}}
}
\rotatebox{270}{
{\includegraphics[width=150pt]{figure_j1650_ratio_topl.ps}}
}
\rotatebox{270}{
{\includegraphics[width=150pt]{figure_cygx1_ratio_topl_nhuntied_noconstraint_obs56.ps}}
}
\rotatebox{270}{
{\includegraphics[width=150pt]{figure_1753xmm_ratio_topl.ps}}
}
\rotatebox{270}{
{\includegraphics[width=150pt]{figure_j1655_ratio_topl.ps}}
}
\rotatebox{270}{
{\includegraphics[width=150pt]{figure_j1118ratio_topl.ps}}
}
\rotatebox{270}{
{\includegraphics[width=150pt]{figure_17497_ratio_topl.ps}}
}
\rotatebox{270}{
{\includegraphics[width=150pt]{figure_j1817_ratio_topl_obs17and18.ps}}
}
\end{center}
\caption{ Data/model ratio to an absorbed powerlaw.  A simple powerlaw
  does not provide a satisfactory model to any of the sources
  above. The excess and curvature below $\approx2$\kev\ is indicative of
  a thermal disk component which is required at over 5$\sigma$ level
  in all sources (see \S3.1). {\it GX339-4:} \epicmos1\ and 2 data are
  shown in black and red respectively. {\it Cygnus X-1:} Obs 1 and 2
  are shown in black and red respectively. {\it J1655-40 and
    J17497-2821:} FI and BI spectrum are shown in black and red
  respectively. {\it J1118+480:} Plus and minus first order spectrum
  are shown in black and red respectively. {\it J1817-330:} Obs 1
  and 2 are shown in black and red respectively. All
  spectra have been re-binned for plotting purposes only. }
\label{fig_ratio_topl}

\end{figure*}

\begin{table*}
\begin{center}
  \caption{Summary of the various black hole binaries physical parameters. }
\label{table_parameters}
\begin{tabular}{lccccccccc}                
  \hline
  \hline
  Source &  Inclination  &  Distance &  Mass &   Total flux$^{ab}$  & Disk flux$^{ab}$  \\
         &  (degrees)    &   (\kpc)  &  ($\msun$) &     &     \\

  GX 339-4 &  10--30 & 6--10& 10--20&  $16.6\pm0.1$&  $2.20^{+0.09}_{-0.08}$     \\
  J1650-500 & 47--70 & 1.9--3.3& 5.3--11.3 & $213.0\pm2$&  $61.2\pm1.2$   \\
  Cygnus X-1 (1)&  25--50 & 2.0--2.2& 7--25& $133.1^{+3.7}_{-3.5}$&  $23.7^{+3.2}_{-3.0}$   \\
  Cygnus X-1 (2)&   25--50 & 2.0--2.2 & 7--25&   $109.6^{+3.7}_{-3.5}$&  $15.5^{+3.2}_{-3.0}$ \\
  J1753.5-0127 &  49--57 & 7.2--10.0& 4--16& $4.10\pm0.01$& $0.13\pm0.01$ \\
  J1655-40 & 68.3--87 & 3.0--3.4& 5.8--6.8  &$9.5\pm0.2$& $0.8\pm0.2$    \\
  J1118+480 & 60--83 & 1.2--2.4& 7--10 &  $13.0\pm0.1$&  $1.11\pm0.05$     \\
  J17497-2821 &  10--80 & 5--10&5--20  &   $15.6^{+1.7}_{-1.4}$&  $6.2^{+1.7}_{-1.4}$    \\
  J1817-330 (1)&  10--80 & 1--15& 4--15 & $7.15\pm0.15$& $2.95^{+0.25}_{-0.27}$     \\
  J1817-330 (2)&   10--80 &  1--15& 4--15  &$1.16^{+0.08}_{-0.07}$& $0.19\pm0.06$   \\
     
\hline
\hline
\end{tabular}

\small Note -- Physical parameters for the various black hole binaries
treated in this work. For references and explanations on the range
quoted above please see \S\ref{observation}. $^a$ Unabsorbed flux in units of $\times10^{-10}$ $\ergpcmsqps$ in the 0.5--10\kev\ energy range  obtained using
the \xspec\ model \cflux\ convolved with the model shown in Table 3. $^b$ The quoted errors for the fluxes are at the 90 per
cent confidence.

\end{center}
\end{table*}

\begin{table*}
  \caption{ Results of fits to simple absorbed powerlaw model.  }

  \begin{tabular}{lccccccccc}                
  \hline
  \hline
  Source &  \nh\ ($\times 10^{22}\pcmsq$)  &  $\Gamma$ &   $N_{\rm PL}$$^c$ &  $\chi^{2}/\nu$\\

GX 339-4 & $0.420\pm0.002$&$1.800\pm0.004$ & $0.256\pm0.001$ &6723.5/1635\\
J1650-500 & $0.549\pm0.002$&$2.562\pm0.006$ & $6.12\pm0.03$ &7937.3/1275  \\
Cygnus X-1 (1)&$0.339\pm0.003$&$1.685\pm0.005$ & $1.66\pm0.01$ &1088.9/724 \\
Cygnus X-1 (2)& $0.346\pm0.004$&$1.669\pm0.006$ & $1.41\pm0.01$ &831.7/685  \\
J1753.5-0127 &$0.175\pm0.001$&$1.666\pm0.003$ &$0.0617\pm0.0002$&3044.6/1499 \\
J1655-40 & $0.525\pm0.003$&$1.660\pm0.005$ &$0.129\pm0.001$&1733.1/1441\\
J1118+480$^a$ & $0.0080^{+0.0001}$&$1.877\pm0.005$ &$0.222\pm0.001$&5451.8/4248 \\
J17497-2821 & $4.349\pm0.003$&$1.50\pm0.01$ &$0.114\pm0.002$&1246.6/1184  \\
J1817-330$^b$ (1)& $0.12(f)$&$3.01\pm0.04$ &$0.215\pm0.004$&403.6/209  \\
J1817-330$^b$ (2)&  $0.12(f)$&$2.01\pm0.07$ &$0.0206\pm0.0001$&107.7/81  \\
\hline
\hline
\end{tabular}

\small Note -- $^a$A lower limit on the absorbing column density  of $8\times10^{19}\pcmsq$ is imposed. $^b$For J1817 \nh\ is frozen at the value indicated. $^c$The powerlaw normalisation is referred as $N$ and is given in units of ${\rm photons{\thinspace keV^{-1}}{\thinspace cm^{-2}}{\thinspace s^{-1}}}$ at 1\kev.  All errors refers to the 90 per cent confidence range for one parameter of interest.

\end{table*}

\begin{table*}
  \caption{Results of fits with the MCD component \diskbb\ and a powerlaw.   }
\label{table_diskbb}
\begin{tabular}{lccccccccc}                
  \hline
  \hline
  Source &  \nh\ ($\times 10^{22}\pcmsq$)   &  $\Gamma$ &   $N_{\rm PL}$$^a$ &  {\it kT} (\kev) & $N_{\rm Diskbb}$ $\times10^3$& Inner radius$^{bc}$ &$\chi^{2}/\nu$\\
  GX 339-4 & $0.495\pm0.006$&$1.665\pm0.009$&$0.222^{+0.003}_{-0.0025}$&$0.254\pm0.006$&$5.02^{+0.80}_{-0.67}$ & $2.7^{+2.1}_{-1.2}$  & 2874.0/1633\\
  J1650-500 &  $0.556\pm0.004$&$2.10\pm0.01$ & $3.40\pm0.06$ & $0.310\pm0.004$&$55\pm4$ & $7.3^{+7.1}_{-3.3}$&1507.9/1273  \\
  Cygnus X-1 (1)& $0.53\pm0.02$&$1.71\pm0.01$ & $1.752\pm0.025$ & $0.194^{+0.005}_{-0.004}$& $236^{+63}_{-54}$ & $5.6^{+7.0}_{-3.0}$&783.1/722   \\
  Cygnus X-1 (2)&  $0.50\pm0.02$&$1.70\pm0.01$ & $1.485\pm0.026$ & $0.194^{+0.007}_{-0.006}$& $155^{+62}_{-49}$ & $4.6^{+6.3}_{-2.7}$&719.2/683   \\
  J1753.5-0127 & $0.197\pm0.004$&$1.61\pm0.01$ &$0.0571\pm0.0006$& $0.274^{+0.015}_{-0.014}$&$0.32^{+0.11}_{-0.08}$ & $1.5^{+2.3}_{-0.9}$&1961.0/1497 \\
  J1655-40 &$0.63\pm0.02$&$1.67\pm0.01$ &$0.1333\pm0.0019$& $0.21\pm0.01$ & $5.4^{+2.7}_{-2.0}$& $5.9^{+6.8}_{-3.3}$& 1618.8/1439 \\
  J1118+480 & $0.022\pm0.003$&$1.69\pm0.01$ &$0.1864\pm0.0028$&$0.21\pm0.01$&$7.4^{+1.4}_{-1.2}$ & $2.3^{+2.6}_{-0.4}$&3747.3/4246 \\
  J17497-2821 & $4.72\pm0.08$&$1.56\pm0.01$ &$0.1282^{+0.0034}_{-0.0032}$&$0.20\pm0.01$&$54^{+49}_{-24}$ & $14^{+40}_{-12}$&1102.5/1182\\
  J1817-330 (1)& 0.12(f)&$2.1\pm0.1$ &$0.097^{+0.013}_{-0.012}$&$0.20\pm0.01$&$27^{+9}_{-6}$ & $14^{+49}_{-13}$&204.0/207  \\
  J1817-330 (2)& 0.12(f)&$1.5\pm0.2$ &$0.012\pm0.003$&$0.21\pm0.04$&$1.3^{+1.5}_{-0.6}$ & $3^{+12}_{-3}$&69.1/79 \\

\hline
\hline
\end{tabular}
\small Note -- Powerlaw nomalisation in units of ${\rm photons{\thinspace keV^{-1}}{\thinspace cm^{-2}}{\thinspace s^{-1}}}$ at 1\kev. $^b$ The errors quoted
for the various inner radii are the 3$\sigma$ errors estimated using
Monte Carlo simulations (see \S3.2). $^c$ The inner radii are given in
units of \rg$=GM/c^2$.

 All errors refers to the 90 per cent confidence range for one parameter of interest.  
\end{table*}

Unfortunately very little is known about the properties of the newly
discovered X-ray binary IGR{\thinspace J17497-2821} (hereafter
  J17497). Paizis et al. (2007) have placed the source in the
  direction of the Galactic centre and argue that it must be either
  very close to or beyond the Galactic centre. We assume the
  distance to J17497 to lie in the range 5--10\kpc. We are not aware
  of any estimates for the mass or inclination which we assume to be
  in the range 5--20$\msun$ and 10--80 degrees respectively.  Paizis
  et al. (2008) presented the broadband \suzaku\ observation of J17497
  in its low-hard state. They find that the data is best modelled with
  the sum of a thermal disk component plus Comptonisation. The
  accretion disk is found to have an inner temperature of
  $\approx$0.25\kev\ and a radius of $\approx$123km for an assumed distance
  of 8\kpc\ and inclination of 60 degrees. For a 10 solar mass black
  hole the authors estimate that the disk extends to $\approx$11\rg\
  after correction for Compton up-scattering.

  We analyse the 2006 September 25 observation (Obs ID 901003010). The
  XIS was operated in the 2x2 and 3x3 editing mode with the ``1/4''
  window employed due to the brightness of the source. The data were
  reduced in the standard way as described above for Cygnus
X-1. Source events were extracted from a circular region 110 pixels
  in radius centred on the source. The background is taken from a
  source free region 110 pixels in radius towards the edge of the
  chip. HXD data is not used in this analysis. Following Paizis et
  al. (2008) we do not use XIS data below 1\kev\ and limit the upper
  energy to 9\kev.

\subsection{J1817-330}

We are not aware of any constraints on the physical parameters of the
newly discovered X-ray binary XTE J1817-330 (hereafter J1817;
Remillard et al. 2006). For this reason we assume the distance, mass
and inclination to be in the range 1--15\kpc, 4--15$\msun$ and 10--80
degrees respectively.

The \xmm\ and \integral\ observations of the source in the soft state
have been reported by Sala et al. (2007). The authors find that the
data can be successfully modelled with the presence of a disk
component with a temperature varying from approximately 1.0\kev\ to
0.7\kev\ and an inner radius of $\approx 50${\thinspace km}, assuming a distance
of 10\kpc\ and 45{\thinspace degrees} inclination. The source was later tracked
through the decline of its outburst by the \swift\ satellite. Rykoff
et al. (2007; hereafter R07) reported on the twenty-one snapshot
observations tracking J1817 from the high-soft to the low-hard
state. They found that the source follows closely the $L_X\propto T^4$
relation during the decline and argued that the accretion disk does
not recede after the state transition down to accretion rates as low
as 0.001$L_{Edd}$. This work was later challenged by Gierlinski et
al. (2008) who, using the same \swift\ data as R07, argued
that irradiation into the disk gives radii which are consistent with
disk truncation.

Here, we re-analyse the data presented in R07 using the same reduction
procedures detailed by the authors. Out of the twenty-two observations
(see Table 1 of R07) thirteen are in the soft state, two are in the
intermediate state and the latter six observations are in the low-hard
state as described in Gierlinski et al. (2008). In this work we will
use observations 17--18 since the data quality in the latter
observations of the low-hard state does not strongly require the
presence of a disk. Following R07, observations 17 and 18 (hereafter
Obs. 1 and 2) were fit in the 0.6--10.0 and 0.3--10.0\kev\
respectively with the absorbing column density fixed at
1.2$\times10^{21}\pcmsq$. We note that the neutral hydrogen column
density in the line of sight to the source is not thought to vary
between observations (Miller, Cackett \& Reis 2009). Table 1 summarises the range
in the various physical parameters used throughout this work.

\section{analysis and results}
\label{analysis}

\begin{table*}
\begin{center}
\caption{Same as Table 3 but for the model containing \ezdiskbb\ plus a powerlaw}

\begin{tabular}{lccccccccc}                
  \hline
  \hline
  Source &  \nh   &  $\Gamma$ &   $N_{\rm PL}$ &  {\it kT} (\kev) & $N_{\rm Ezdiskbb} \times 10^3$&  Inner radius &$\chi^{2}/\nu$\\

  GX 339-4 & $0.495^{+0.006}_{-0.005}$&$1.666^{+0.009}_{-0.008}$&$0.222^{+0.003}_{-0.002}$&$0.241\pm0.005$&$1.1^{+0.2}_{-0.1}$&$3.5^{+4.4}_{-2.2}$ &2885.5/1633\\
  J1650-500 &  $0.556\pm0.004$&$2.10\pm0.01$ & $3.42\pm0.06$ & $0.293\pm0.004$&$12.0^{+0.9}_{-0.8}$ &$9^{+14}_{-6}$&1509.1/1273  \\
  Cygnus X-1 (1)&$0.52\pm0.02$&$1.71\pm0.01$ & $1.751\pm0.025$&$0.185^{+0.005}_{-0.004}$& $49^{+13}_{-11}$ &$7^{+13}_{-5}$&782.4/722   \\
  Cygnus X-1 (2)&  $0.50\pm0.02$&$1.70\pm0.01$ & $1.484\pm0.026$ & $0.185\pm^{+0.007}_{-0.006}$& $32^{+13}_{-10}$ &$6^{+11}_{-4}$&718.9/683  \\
  J1753.5-0127 & $0.197\pm0.004$&$1.609\pm0.007$ &$0.0572\pm0.0006$& $0.26\pm0.01$&$0.075\pm0.02$ &$2.0^{+4.1}_{-1.4}$&1969.2/1497 \\
  J1655-40 &$0.63\pm0.02$&$1.67\pm0.01$ &$0.133\pm0.002$& $0.20\pm0.01$&$1.1^{+0.6}_{-0.4}$&$7.4^{+12.0}_{-4.9}$& 1618.9/1439 \\
  J1118+480 & $0.022\pm0.003$&$1.69\pm0.01$ &$0.186\pm0.003$&$0.20\pm0.01$&$1.6^{+0.3}_{-0.2}$&$3.0^{+4.6}_{-1.0}$ &3746.0/4246 \\
  J17497-2821 & $4.72\pm0.08$&$1.56\pm0.01$ &$0.128\pm0.003$&$0.19\pm0.02$&$10.7^{+9.4}_{-4.7}$ &$17^{+57}_{-15}$&1102.4/1182\\
  J1817-330 (1)&  0.12(f)&$2.1\pm0.1$ &$0.098^{+0.013}_{-0.012}$&$0.19\pm0.01$&$5.9^{+1.9}_{-1.3}$ &$18^{+72}_{-17}$&204.5/207    \\
  J1817-330 (2)&   0.12(f)&$1.5\pm0.2$ &$0.012\pm0.003$&$0.20^{+0.04}_{-0.03}$&$0.27^{+0.32}_{-0.13}$ &$4^{+17}_{-3}$&69.1/79  \\
     
\hline
\hline
\end{tabular}
\end{center} 

\end{table*}

The X-ray spectrum (0.1--100\kev) of stellar mass black hole binaries
in the low-hard state above a few \kev\ is usually thought to be
dominated by inverse Compton scattering of the soft thermal-disc
photons in a cloud of hot electrons or ``corona'' (Zdziarski \&
Gierlinski 2004). The geometry of the Comptonisation region is still
not firmly established. Certain models have hot electrons confined in
magnetic flares above the cool accretion disk (Beloborodov 1999;
Merloni, Di Matteo \& Fabian 2000; Merloni \& Fabian 2001) whilst
others predict that the hard X-ray emission originates in the base of
a jet (Markoff \& Nowak 2004; Markoff, Nowak \& Wilms 2005). A further
possibility is that the accretion flow consists of a thin disk
truncated at large distances from the black hole ($R_{tr}\ga10^3$\rg;
Esin et al. 1997) and the hard emission originates in the hot,
quasi-spherical advection-dominated accretion region close to the
central source. In all these cases the X-ray continuum in the
$\sim$2--10\kev\ energy range can be phenomenologically modelled by a
simple powerlaw with a photon index between approximately 1.4--2, with major differences between the models arising at
energies greater than $\sim$30 and below $\sim$2\kev.

\begin{figure}
\rotatebox{270}{
{\includegraphics[width=155pt]{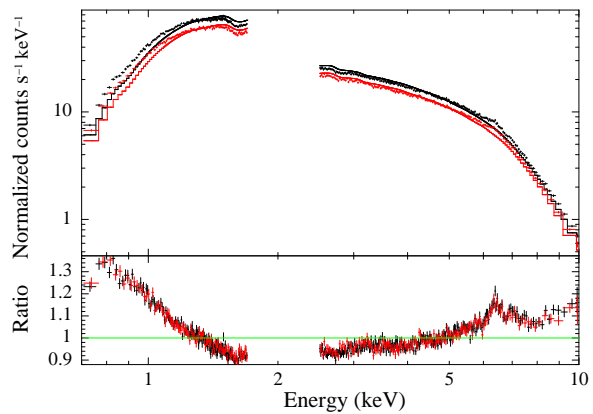}}
}
\caption{ Ratio to powerlaw for Cyg{\thinspace X-1} when the column density \nh\
  is constrained to 5--8$\times10^{21}\pcmsq$. See \S3.}
\label{cygx1_nhconstrained}
\end{figure}

We begin our analyses by fitting the various spectra up to a maximum
energy of 10\kev\ with a powerlaw modified by absorption in the
interstellar medium (\phabs\ model in \xspec). The equivalent neutral
hydrogen column density (\nh) is initially a free parameter in all
observations. It is clear from Table 2 and Fig.\ref{fig_ratio_topl}
that the model is a poor description of the data to the majority of
the objects. Strong curvature below $\sim$2\kev\ is clearly seen in
all sources other than J17497 and J1817 possibly due to the poorer
low-energy coverage in those observations. The excess and curvature
seen at soft energies is characteristic of a thermal disc
emission and {\it cannot} be properly modelled by variation in the
absorbing column density. This is a consequence of the high quality
data presented here. As an example of the way artificial variation in
the absorbing column density can act to mask out the presence of a
disk component we constrain the value of \nh\ for Cyg{\thinspace X-1} to
5--9{\thinspace $\times10^{21}\pcmsq$}; this covers the largest published range of
\nh\ for this object (Dotani et al. 1997; Takahashi et al. 2001;
Miller et al. 2002; Torres et al. 2005; Makishima et al. 2008). Fitting
the spectra with a powerlaw now results in a much more distinct soft
excess as can be seen in Fig. \ref{cygx1_nhconstrained}. A change in
\nh\ from $\approx$3.4$\times 10^{21}\pcmsq$ (Table 2) to $5\times
10^{21}\pcmsq$ results in a dramatic difference to the modelled
spectra of Cyg{\thinspace X-1} in the low-hard state. A similar behaviour is seen
in J1118 where \nh\ peaks at the lower limit imposed of $ 8\times
10^{19}\pcmsq$.  This behaviour has been reported for Cyg{\thinspace X-1} by
Takahashi et al. (2001) using \asca\ data and possibly explains the
lower value of \nh\ found in the single \swift\ pointing of J1655 in
the low-hard state (see Tables 4 and 5 in Brocksopp et al. 2006).

Evidence of reflection features in the form of an excess at approximately
4--7\kev\ can be seen in all spectra other than J1118, J17497
and J1817. In the work that follows we will ignore the 4--7\kev\ range
in GX 339-4 and J1650 and the 5--7\kev\ range in Cyg{\thinspace X-1}, J1753 and
J1655. The reflection features will be individually explored in
\S\ref{innerrad_reflection}. Table 2 details the parameters obtained
when fitting the various spectra with a simple absorbed
powerlaw. Although none of the fits are statistically acceptable we
show the results obtained with this model to emphasise the
necessity of a further disk component in the spectra of black hole binaries
in the low-hard state.

\subsection{The requirement for a MCD component}
\label{analyses_mcd}

The most widely used multicolour disk model (MCD) for the study of
black holes accretion is the \diskbb\ model (Mitsuda et
al. 1984). This simple model describes the spectrum from a
geometrically thin and optically thick (Shakura \& Sunyaev 1973)
accretion disk consisting of multiple blackbody components. It is
parametrised by the colour temperature and a normalisation factor
defined as $(r/d)^2cos\theta$ where $r$ is the inner radius of the
accretion disk in km, $d$ is the distance to the source in units of
10\kpc\ and $\theta$ is the disk inclination. Table 3 details the
parameters obtained when \diskbb\ is used to model the soft disk
emission in the various sources. The addition of this model is
required at greater than the 5$\sigma$ level in {\it all} observations
established using the \ftest\ command in \xspec. The low temperature
found in all sources would render the strong detection of the disk
component impossible with \rxte.

The multicolour disk model \diskbb\ assumes a non-zero torque boundary
condition at the inner edge of the accretion disk (Gierlinski et
al. 1999). This is contrary to the standard idea where the accretion
disk extends down to an inner radius \rin\ at which point the material
free-falls rapidly with negligible viscous interaction.  In order to
test the effect of zero-torque and non-zero torque boundary
conditions, Zimmerman et al. (2005) constructed the MCD model
\ezdiskbb\ implementing the zero-torque boundary condition. Similarly
to the \diskbb\ model, \ezdiskbb\ has as a parameter the colour
temperature as well as a normalisation described as
$(1/f)^4(r/d)^2cos\theta$, where we remind the reader that the colour
correction factor, $f$ is the ratio between the colour temperature and
the blackbody temperature of the disk, and all other symbols are
similar to that of \diskbb.

To confirm that the results presented above do not depend on the
choice of MCD model used, we have re-analysed the data with the model
\ezdiskbb. In what follows we assume the standard value of 1.7 for
the colour correction factor (Shimura \& Takahara 1995) and set a
lower and upper limit of 1.3 and 2 respectively when calculating
errors on the derived inner radius (see \S\ref{innerrad_diskpn}). Table 4 shows the results of the fits with the \diskbb\ model replaced by
the zero-torque model \ezdiskbb. Statistically we find no difference
in the models between most of the fits, with the two models also
giving very similar colour temperatures. We note here that all sources
in our sample strongly require the presence of a thermal
component. The question we will address below is whether this thermal
emission arises from a small region extending close to the ISCO or
from a larger, cooler region truncated far from the black hole.

\begin{figure}
\rotatebox{270}{
{\includegraphics[width=165pt]{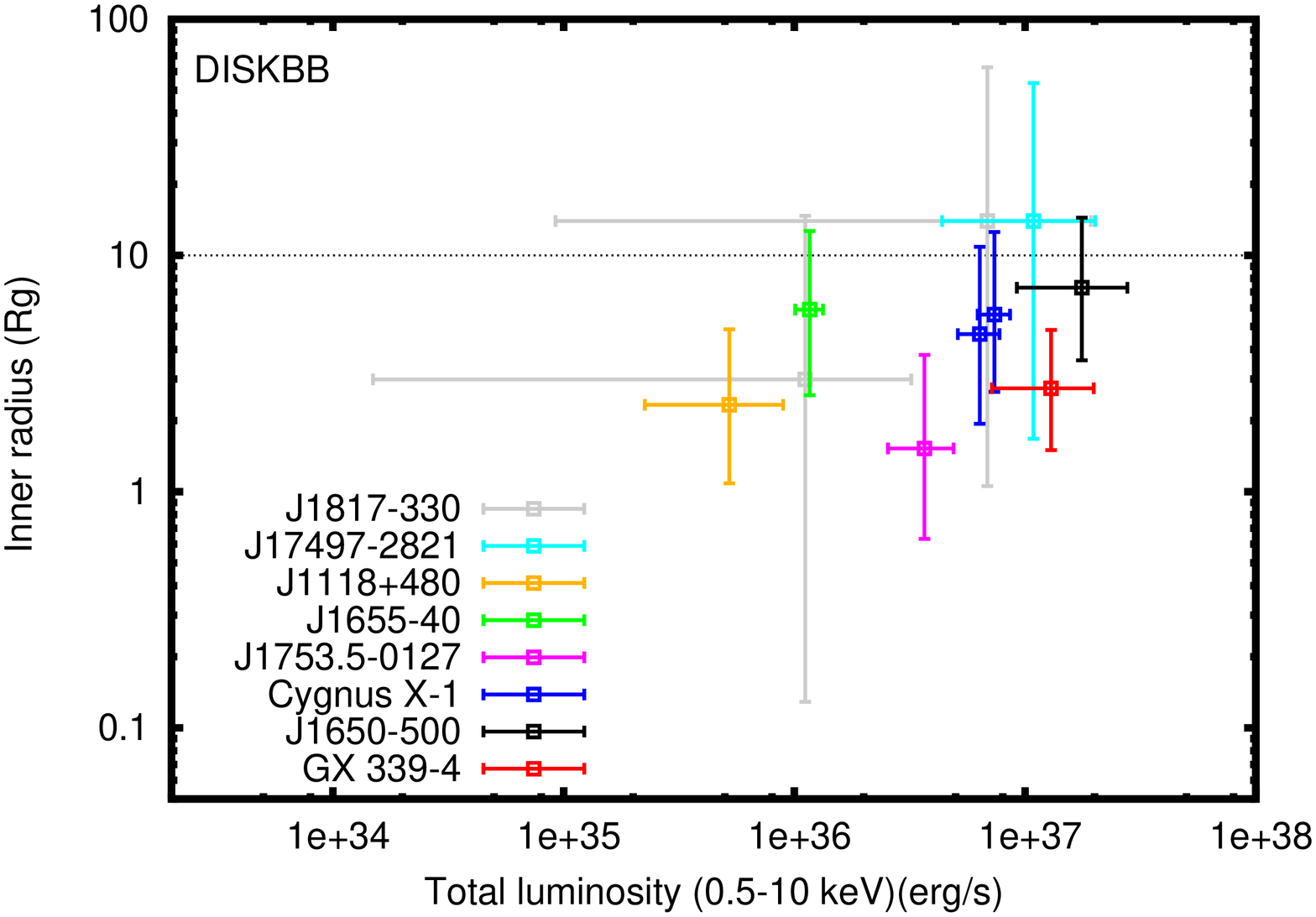}}
}
\rotatebox{270}{
{\includegraphics[width=165pt]{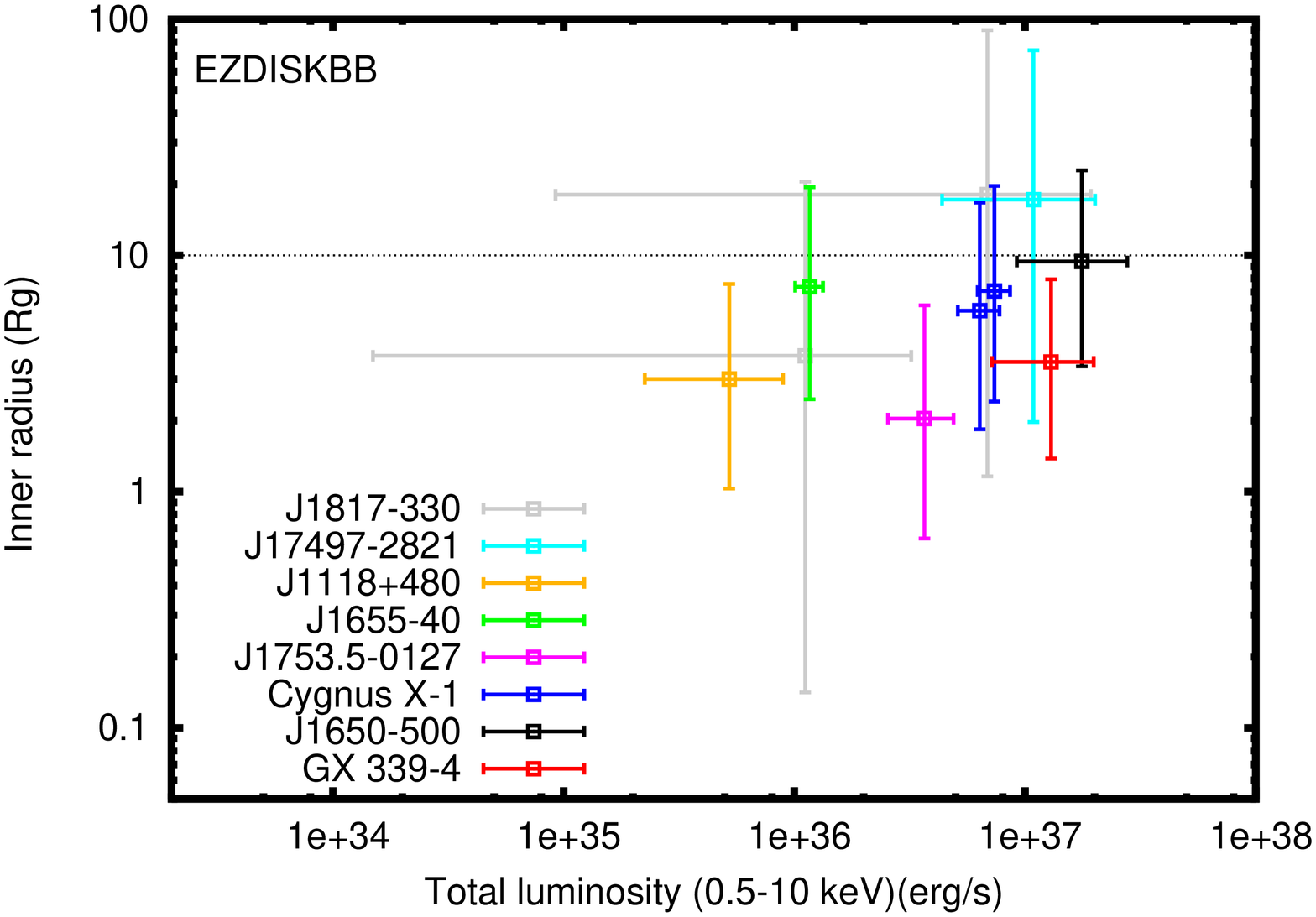}}
}
\caption{ Inner radius versus Total (unabsorbed) luminosity
  (0.5--10\kev). {\it (Top:}) \rin\ derived from the normalisation of
  the \diskbb\ MCD model. {\it (Bottom:}) Similar to above but for
  the \ezdiskbb\ MCD model. The horizontal dashed line shows
  \rin=10\rg. The errors represent the 3$\sigma$ confidence level
   estimated using Monte Carlo simulations (see \S 3.2).}
\label{fig_rin_lumin}

\end{figure} 
\subsection{Inner radius from the thermal component}
\label{innerrad_diskbb}

The observed thermal disk flux depends on the emitting area
($\propto$\thinspace\rin$^2$), inclination and distance to the
source. These are parametrised in the normalisation of the \diskbb\
and \ezdiskbb\ component as described above. Using the constraints on
the physical parameters of the various sources along with the measured
normalisations we derived constraints on the inner-most radius of the
disk for the various sources. The derived values for the inner-radius of the various sources in units of gravitational radii are shown in Tables 3 and 4.

\begin{figure}
\rotatebox{270}{
{\includegraphics[width=165pt]{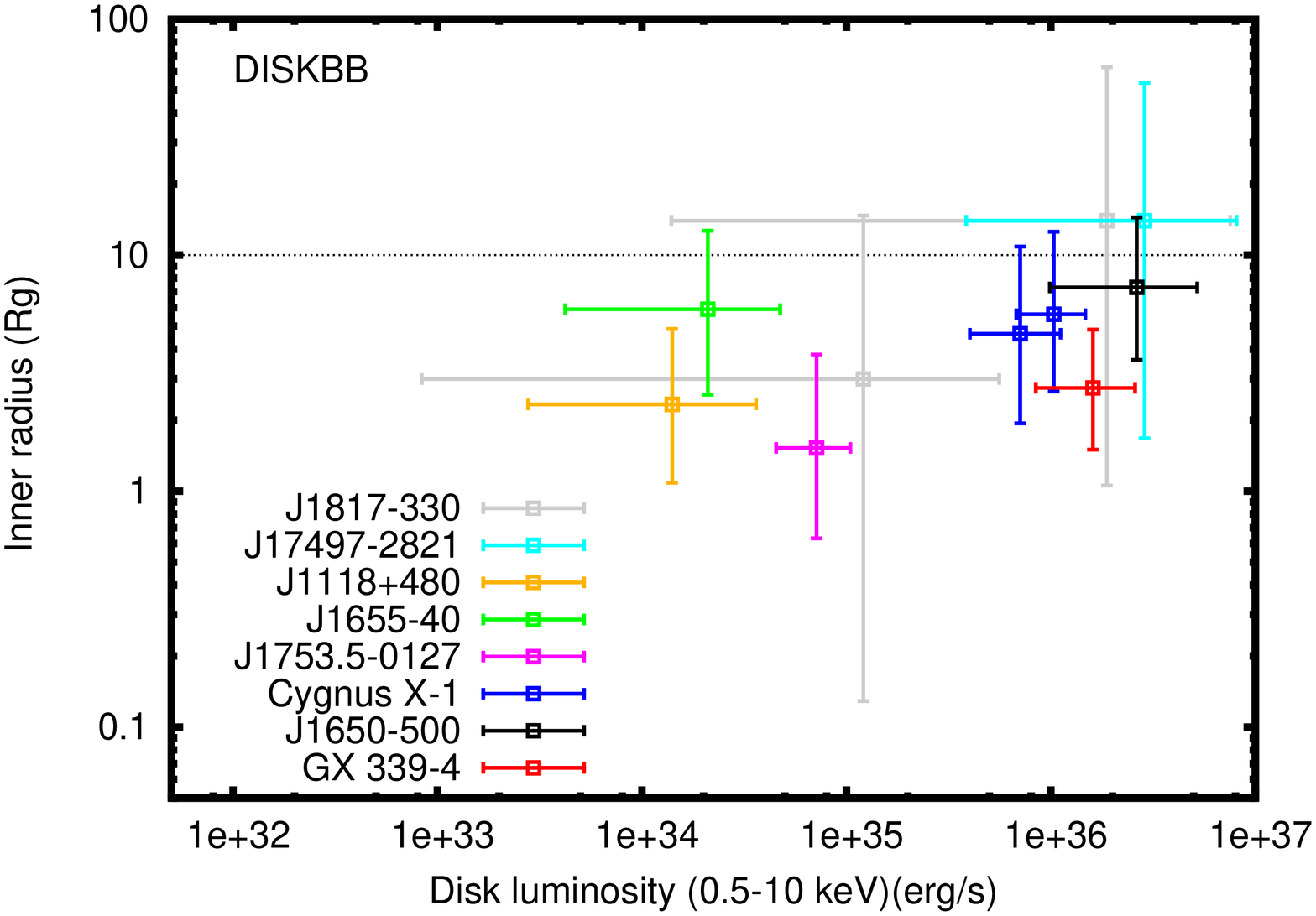}}
}
\caption{ Inner radius versus disk luminosity (0.5--10\kev). The inner
  radius, \rin\ is derived from the normalisation of the \diskbb\ MCD
  model. The horizontal dashed line shows \rin=10\rg. All errors
  represent the 3$\sigma$ confidence level. }
\label{fig_rin_disklumin}

\rotatebox{270}{
{\includegraphics[width=165pt]{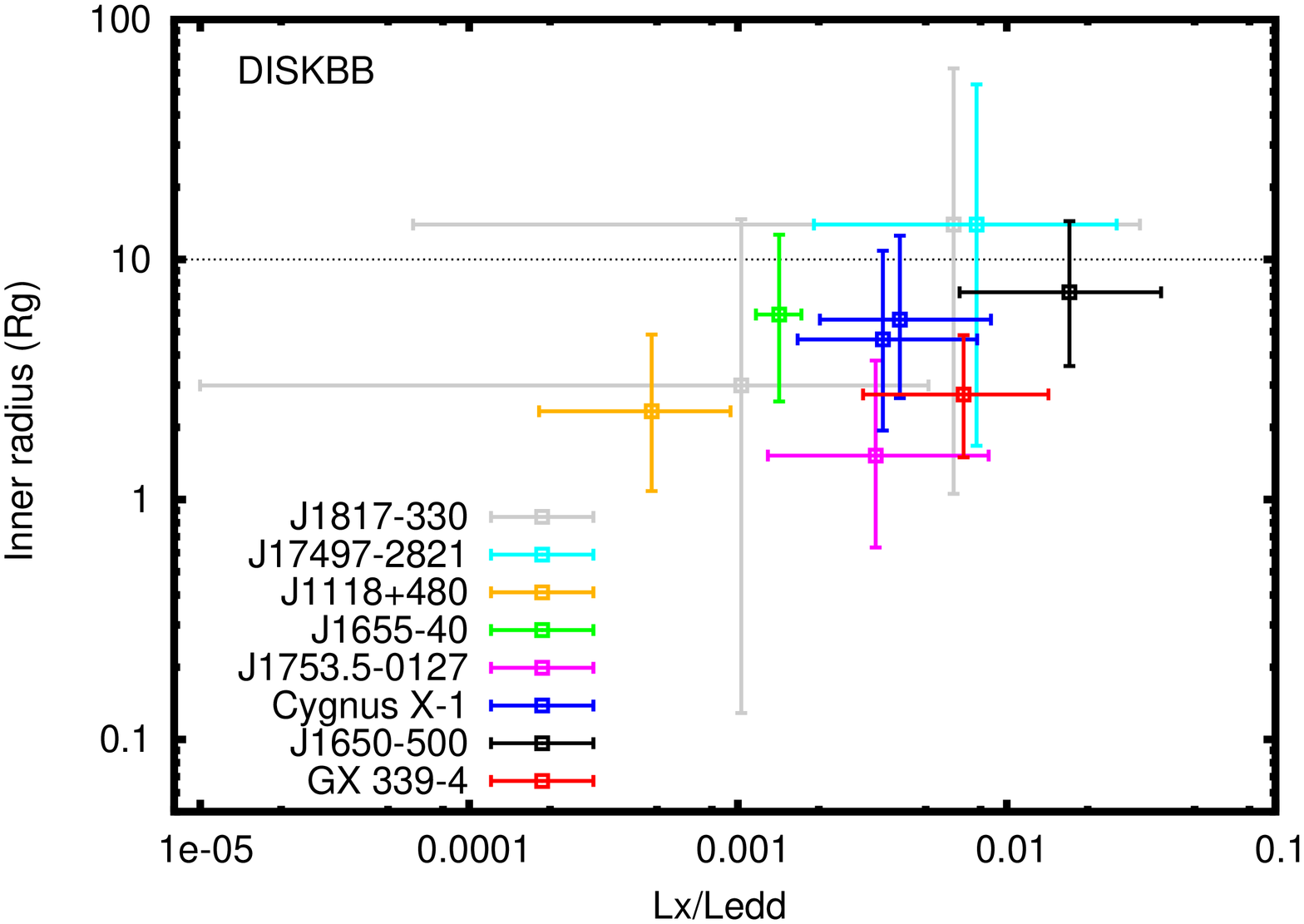}}
}
\rotatebox{270}{
{\includegraphics[width=165pt]{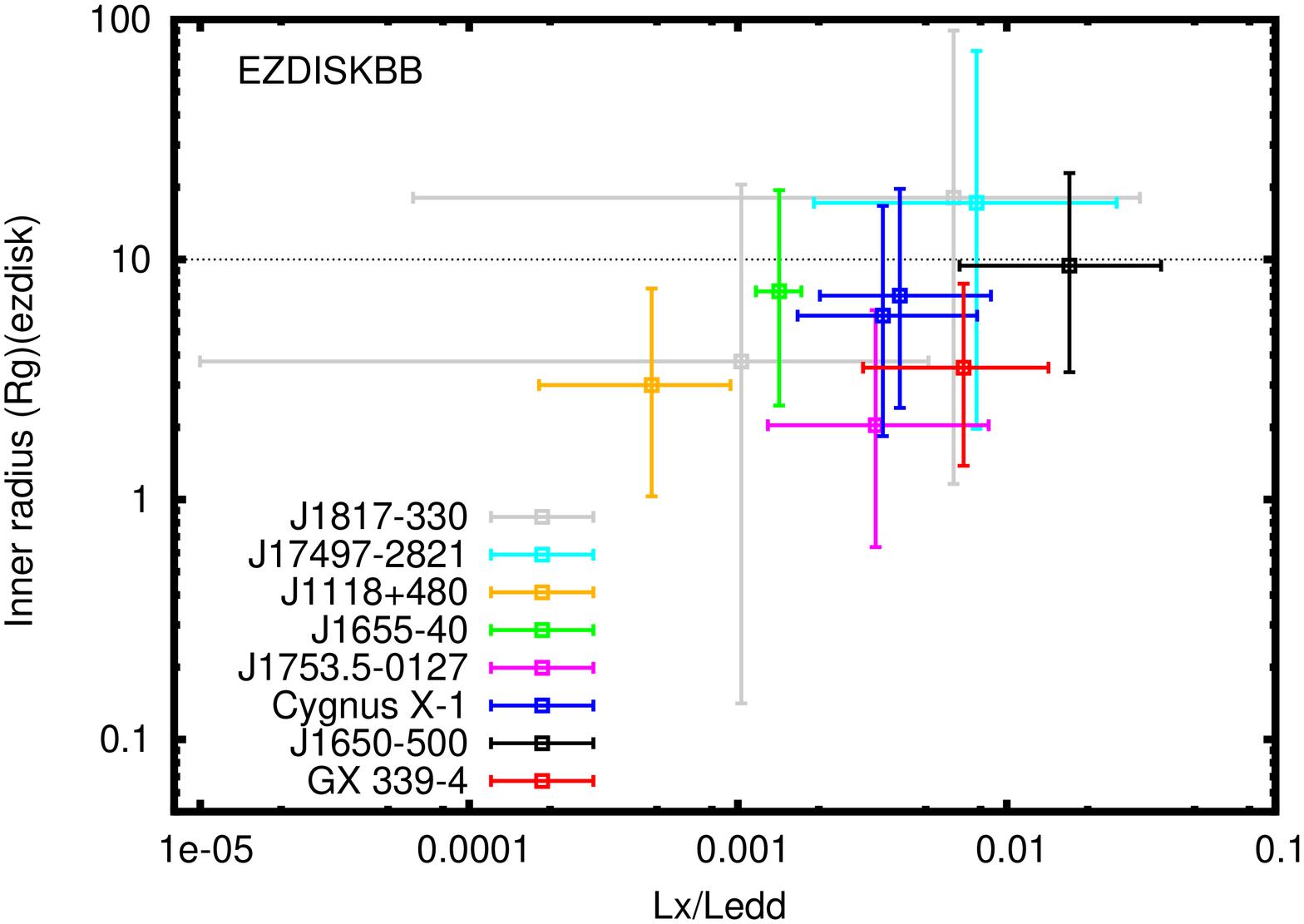}}
}
\caption{ Inner radius versus $L_X/L_{Edd.}$ {\it (Top:}) \rin\ from \diskbb. {\it (Bottom:}) \rin\ from \ezdiskbb. }
\label{fig_rin_loverled}
\end{figure}

\begin{figure}

\rotatebox{270}{
{\includegraphics[width=165pt]{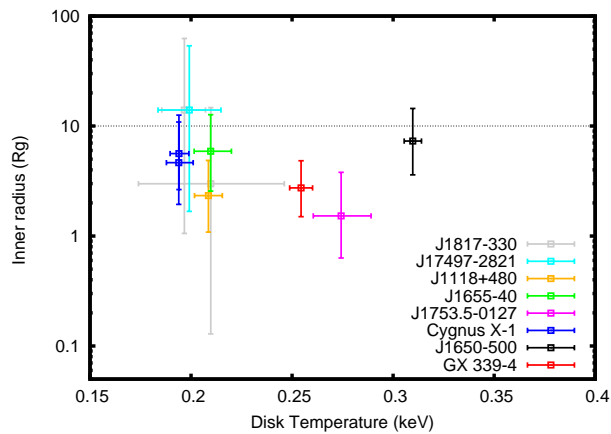}}
}
\caption{ Inner radius obtained from the normalisation of the \diskbb\
  model versus disk temperature. The errors on the inner radii and
  disk temperature are 3 and 1.64$\sigma$ respectively.  }
\label{fig_rin_kt}

\end{figure}

\begin{figure}

\rotatebox{270}{
{\includegraphics[width=165pt]{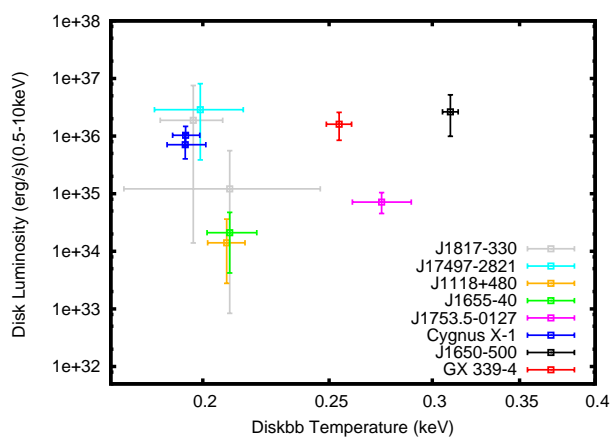}}
}

\caption{ Disk luminosity as a function of disk temperature. Assuming
  a luminosity-temperature relation of $L_{disk} \propto T_{BB}^4$ we
  would expect to see a factor of 2--3 change in temperature for the
  range in luminosity shown in the plot. The errors on the luminosity and
  disk temperature are 3 and 1.64$\sigma$ respectively.   }

\end{figure}

Figure 3 shows the derived inner radii from the normalisations of
the \diskbb\ (Top) and \ezdiskbb\ (Bottom) model as a function of the
total unabsorbed luminosity. The inner radii obtained from the
\diskbb\ model are shown as a function of disk luminosity in
Fig. 4. The 3$\sigma$ errors on the inner radii and luminosities shown
in Figs. 3--6 were estimated using Monte Carlo simulations. A uniform
distribution was assumed for the masses, distances and inclinations of
the various sources with the range shown in Table{\thinspace 1}. The
normalisation of the MCD components and the fluxes were assumed to
have a Gaussian distribution.  When constraints, however mild, on the
inclinations, masses and distances are available we see from Figs. 3
and 4 that the derived inner radii from the thermal continuum using
the model \diskbb\ are below $\sim$10\rg\ and {\it always} consistent
with extending to the innermost stable circular orbit. In all cases
other than the two sources with little constraints on the physical
parameters (J17497 and J1817) the maximum value possible for the inner
radius lies below $\sim$20\rg\ at the 3$\sigma$ confidence level. The
results obtained from the \ezdiskbb\ model (Fig. 3; Bottom) are
similar to that of \diskbb\ but for a factor of $\approx$1.34
higher. Figs. 5 and 6 show the inner radii as a function of Eddington
ratio and colour temperature respectively.

It can be seen from Figs. 3 to 5 that the various black hole binaries
investigated  cover a wide range of both luminosities and Eddington
ratios, with the least luminous source (J1118) being approximately two
orders of magnitude less luminous than J1650. It is interesting to
note from Fig. 5 that the majority of sources are clustered at
$\sim$5--9$\times10^{-3}L_{Edd}$ which is indicative of a mild
selection effect in the sample.

\begin{table*}
\begin{center}
\begin{tabular}{lccccccccc}                
  \hline
  \hline
  Source &  \nh  &  $\Gamma$ &   $N_{\rm PL}$ &  {\it kT} (\kev) & $N_{\rm 6}$&   $\chi^{2}/\nu$\\
GX 339-4 &  $0.496^{+0.006}_{-0.005}$&$1.666^{+0.009}_{-0.008}$&$0.222^{+0.003}_{-0.002}$&$0.241\pm0.005$&$0.09\pm0.01$&2887.6/1633 \\
J1650-500 &  $0.556\pm0.004$&$2.10\pm0.01$ & $3.42\pm0.06$ & $0.293\pm0.004$&$1.03^{+0.08}_{-0.07}$ &1508.9/1273 \\
Cygnus X-1 (1)& $0.52\pm0.02$&$1.71\pm0.01$ & $1.75\pm0.02$&$0.185^{+0.005}_{-0.004}$& $4.1^{+1.1}_{-0.9}$ &782.5/722 \\
Cygnus X-1 (2)& $0.50\pm0.02$&$1.70\pm0.01$ & $1.48\pm0.03$&$0.185^{+0.007}_{-0.006}$& $2.71^{+1.1}_{-0.9}$ &718.9/683 \\
J1753.5-0127 &  $0.198^{+0.004}_{-0.005}$&$1.610^{+0.006}_{-0.008}$ &$0.0573^{+0.0005}_{-0.0006}$& $0.25\pm0.01$&$0.006\pm0.002$ &1970.4/1497\\
J1655-40 & $0.63\pm0.02$&$1.67\pm0.01$ &$0.133\pm0.002$& $0.20\pm0.01$&$0.096^{+0.047}_{-0.036}$& 1618.9/1439 \\
J1118+480 & $0.023\pm0.003$&$1.69\pm0.01$ &$0.186\pm0.003$&$0.199^{+0.006}_{-0.007}$&$0.135^{+0.03}_{-0.02}$ &3744.9/4246 \\
J17497-2821 & $4.72\pm0.08$&$1.56\pm0.01$ &$0.128\pm0.003$&$0.191\pm0.015$&$0.9^{+0.8}_{-0.4}$ &1102.4/1182\\
J1817-330 (1)&  0.12(f)&$2.1\pm0.1$ &$0.098^{+0.013}_{-0.012}$&$0.19\pm0.01$&$0.5^{+0.2}_{-0.1}$ &204.5/207    \\
J1817-330 (2)&  0.12(f)&$1.5\pm0.2$ &$0.012\pm0.003$&$0.20^{+0.04}_{-0.03}$&$0.02^{+0.03}_{-0.01}$ &  69.0/79\\
\

\end{tabular}

\end{center}

\end{table*}

\begin{table*}
\begin{center}
\begin{tabular}{lccccccccc} 
\

 Source &   \nh  &  $\Gamma$ &   $N_{\rm PL}$ &  {\it kT} (\kev) & $N_{\rm  100}$&  $\chi^{2}/\nu$\\
GX 339-4 & $0.499^{+0.004}_{-0.008}$&$1.67\pm0.01$&$0.222^{+0.002}_{-0.003}$&$0.254^{+0.008}_{-0.004}$&$0.00215^{+0.0002}_{-0.0004}$&2880.4/1633 \\
J1650-500 & $0.557\pm0.004$&$2.10\pm0.01$ & $3.40\pm0.06$ & $0.311\pm0.004$&$0.022\pm0.002$ &1506.7/1273  \\
Cygnus X-1 (1)& $0.53\pm0.02$&$1.71\pm0.01$ & $1.75\pm0.02$&$0.195^{+0.005}_{-0.004}$& $0.10^{+0.03}_{-0.02}$ &783.42/722   \\
Cygnus X-1 (2)& $0.50\pm0.02$&$1.70\pm0.01$ & $1.49\pm0.03$&$0.195^{+0.007}_{-0.006}$& $0.06^{+0.03}_{-0.02}$ &719.5/683  \\
J1753.5-0127 &  $0.196\pm0.002$&$1.61\pm0.01$ &$0.0570^{+0.0007}_{-0.0004}$& $0.28^{0.01}_{-0.02}$&$0.00012^{+0.00005}_{-0.00002}$ &1963.5/1497\\
J1655-40 &  $0.63\pm0.02$&$1.67\pm0.01$ &$0.133\pm0.002$& $0.21\pm0.01$&$0.002\pm0.001$& 1618.6/1439\\
J1118+480 & $0.023\pm0.003$&$1.69\pm0.01$ &$0.186\pm0.003$&$0.210\pm0.007$&$0.0030^{+0.0006}_{-0.0005}$ &3742.7/4246 \\
J17497-2821 &  $4.72\pm0.08$&$1.56\pm0.01$ &$0.128\pm0.003$&$0.20\pm0.02$&$0.02^{+0.02}_{-0.01}$ &1102.5/1182\\
J1817-330 (1)&  0.12(f)&$2.1\pm0.1$ &$0.097^{+0.013}_{-0.012}$&$0.20\pm0.01$&$0.011^{+0.003}_{-0.002}$ & 203.6/207 \\
J1817-330 (2)&  0.12(f)&$1.5\pm0.2$ &$0.012\pm0.003$&$0.21\pm0.04$&$0.0005^{+0.0006}_{-0.0002}$ & 69.0/79 \\

\hline
\hline
\end{tabular}
\end{center}

\caption{Results for \diskpn\ fits with the inner radius fixed at 6\rg\ (Top) and 100\rg\ (Bottom). It can be seen that for most sources the parameters and statistics do not vary between models. The only exception to this are the large differences in the normalisation of \diskpn. We can use this difference in conjunction with physical parameters (inclination, mass and distances; see Table \ref{table_parameters}) to differentiate between these two contradictory interpretations (see \S 3.4). All errors are quoted at the 90 per cent confidence level.}

\end{table*}

\subsection {Standard disk in the low-hard state}

The standard Shakura-Sunyaev (SS) model predicts the accretion disk
temperature to be $\sim(M/10\msun)^{-1/4}(L/L_{Edd})^{1/4}$\kev\
(Shakura \& Sunyaev 1973) for a disk extending to the ISCO. This gives
a range in temperature of $\approx$0.15--0.35\kev\ for the observed
values of $L/L_{Edd}$ ($\sim$$5\times10^{-4}$ to $1.5\times10^{-2}$;
Fig. \ref{fig_rin_loverled}).  It can immediately be seen from Figs. 6
and 7 that the range in temperature observed is in agreement with that
predicted by the SS model given the values of Eddington ratio found
here. This shows that the behaviour of the accretion disk in the
low-hard state for the various sources observed are characteristic of
that predicted for a standard, geometrically thin disk. The results
presented so far agree with that presented by Rykoff et al. (2007)
where the authors claim to see the presence of a standard accretion
disk extending down to the ISCO until at ``least'' $10^{-3}L_{Edd}$ in
J1817.

\begin{figure*}
\rotatebox{0}{
{\includegraphics[width=225pt, clip]{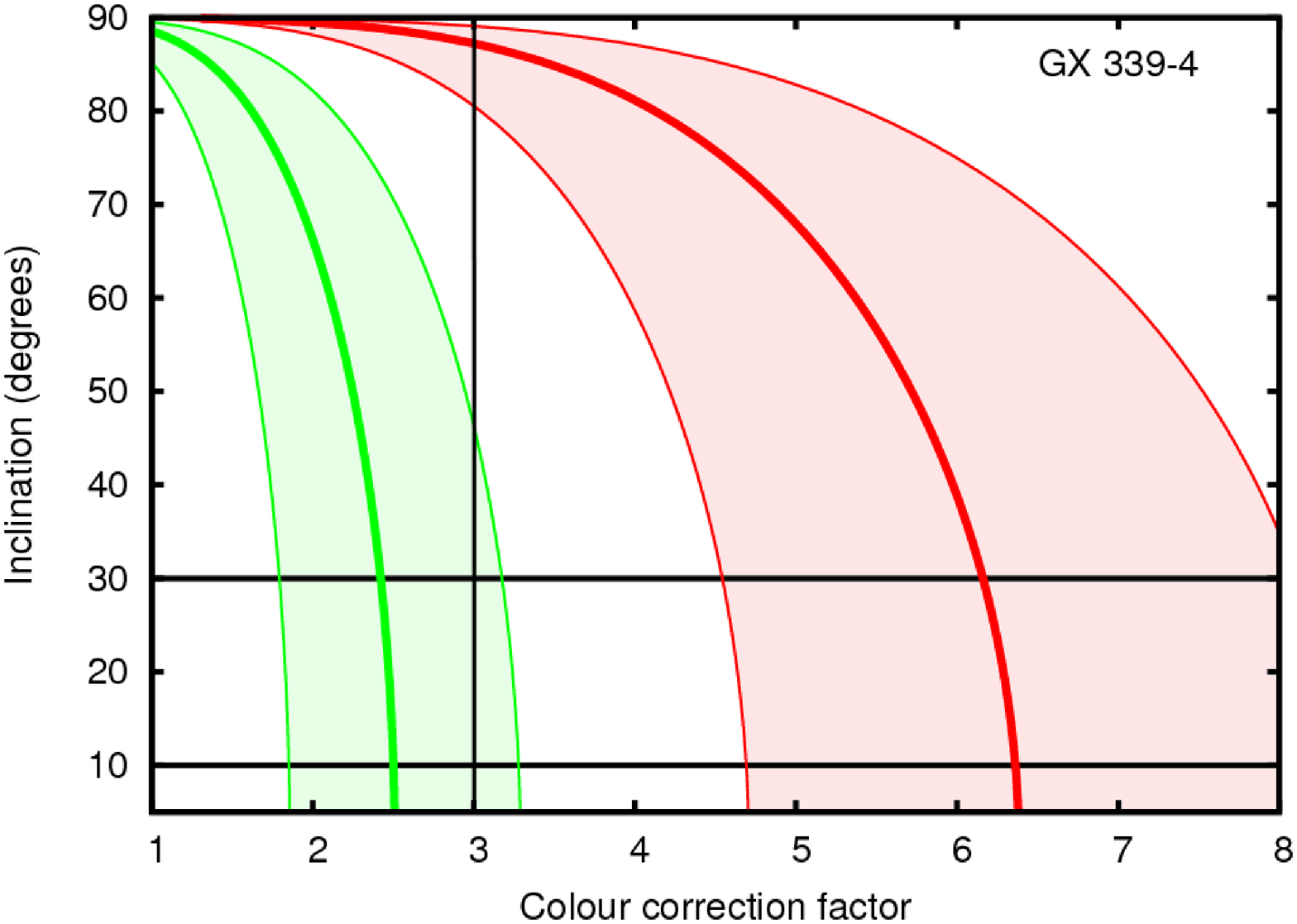}}
}
\rotatebox{0}{
{\includegraphics[width=225pt,clip]{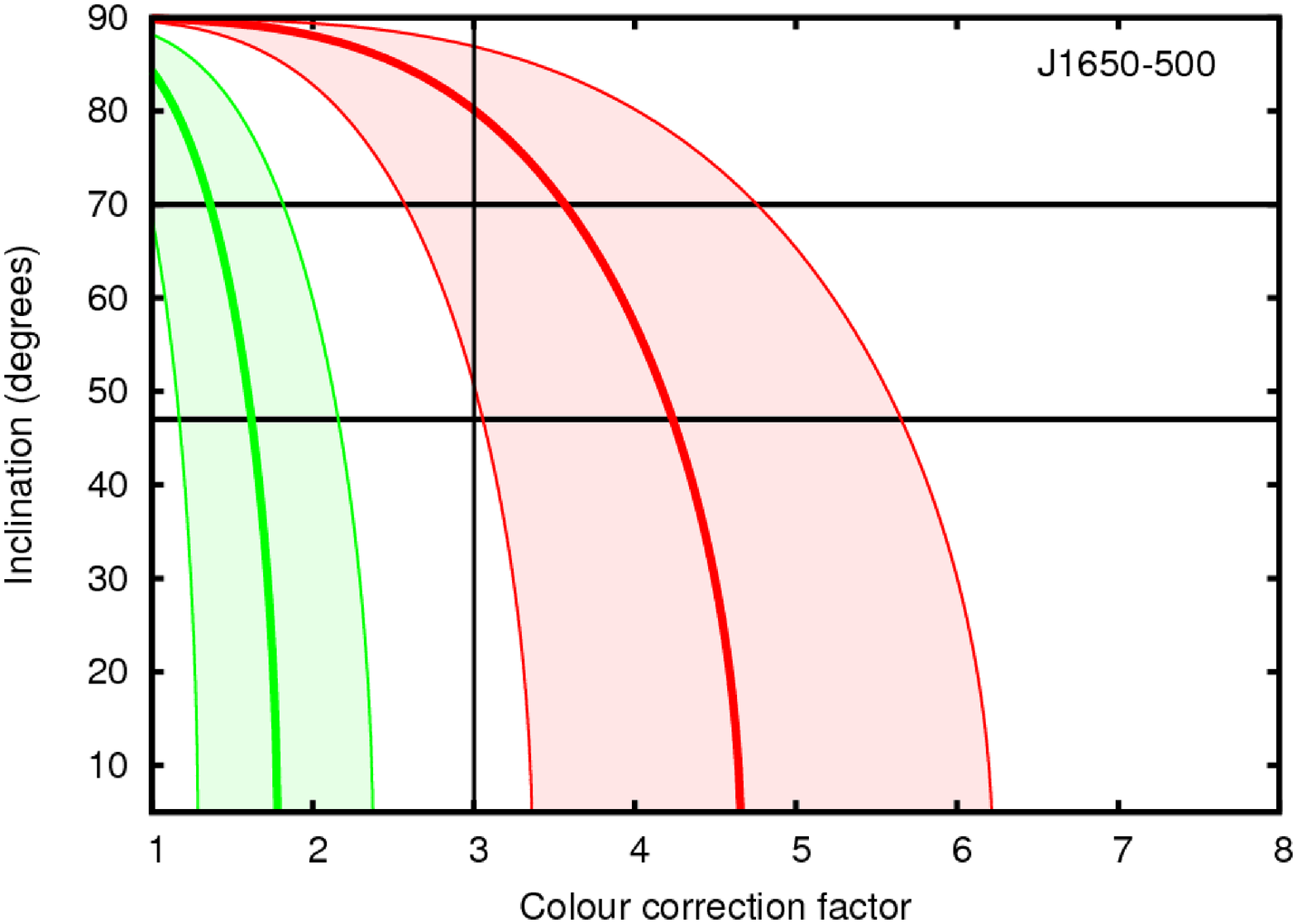}}
}
\rotatebox{0}{
{\includegraphics[width=225pt,clip]{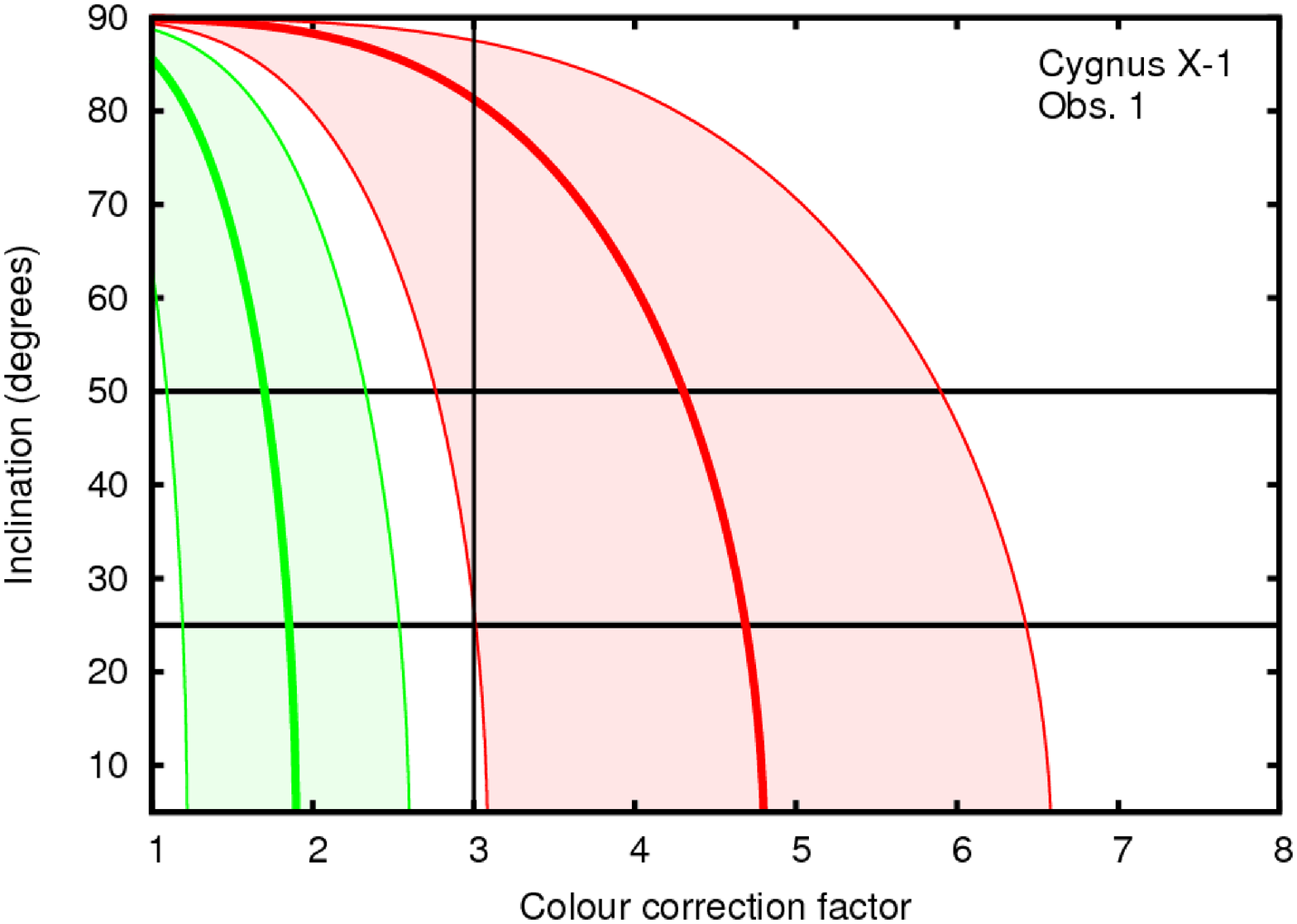}} 
}
\rotatebox{0}{
{\includegraphics[width=225pt,clip ]{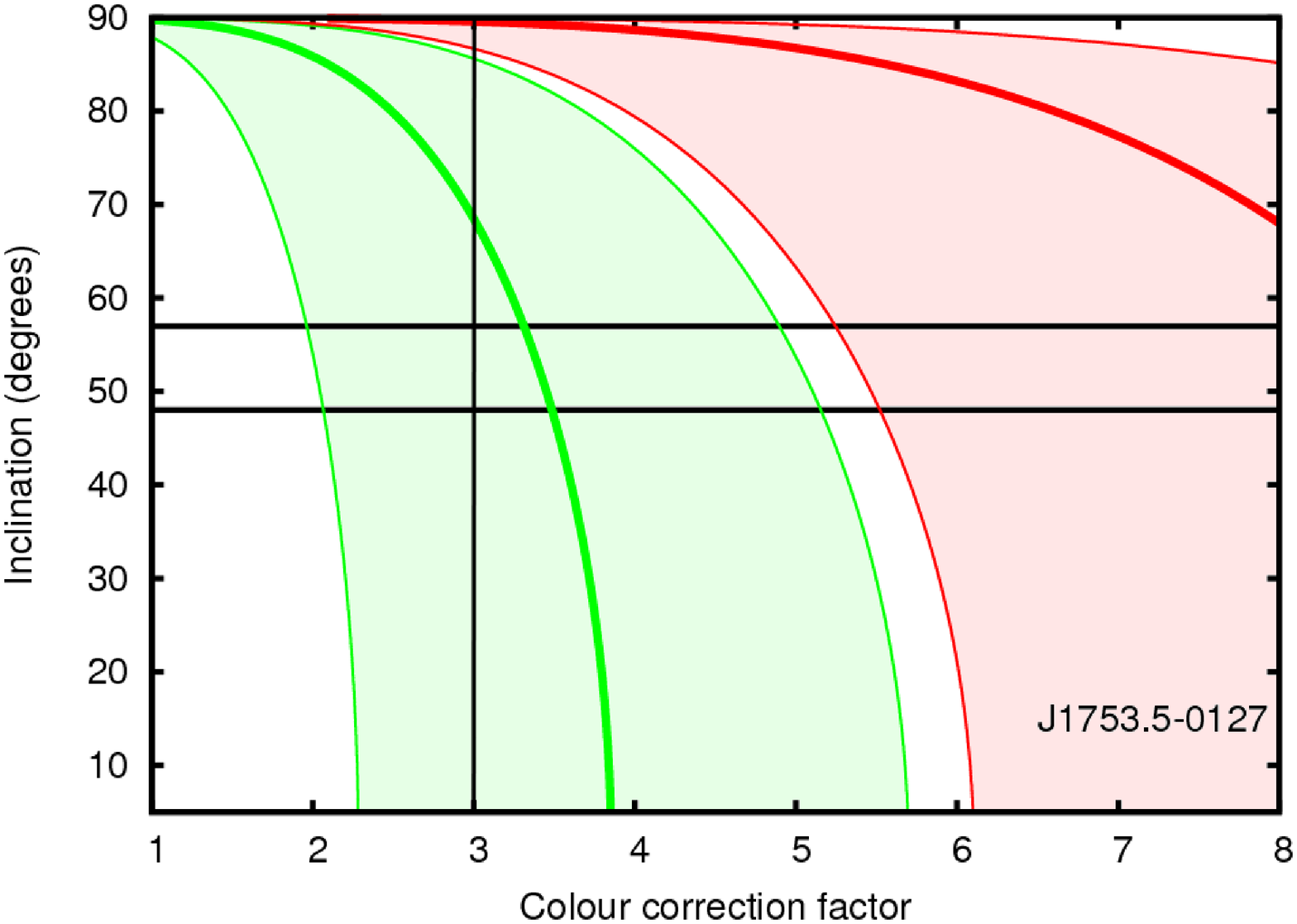}}
}
\rotatebox{0}{
{\includegraphics[width=225pt,clip ]{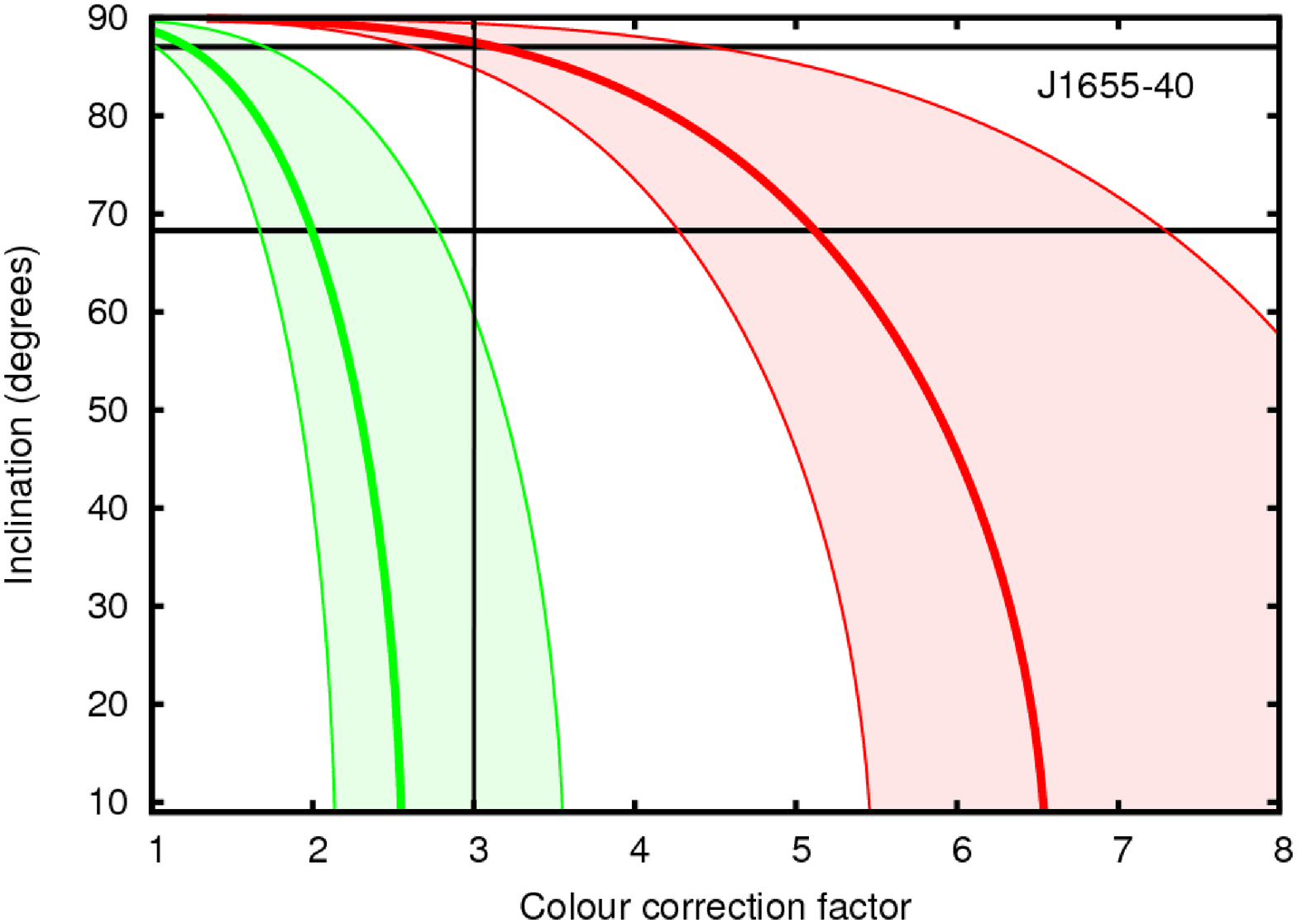}}
}
\rotatebox{0}{
{\includegraphics[width=225pt,clip ]{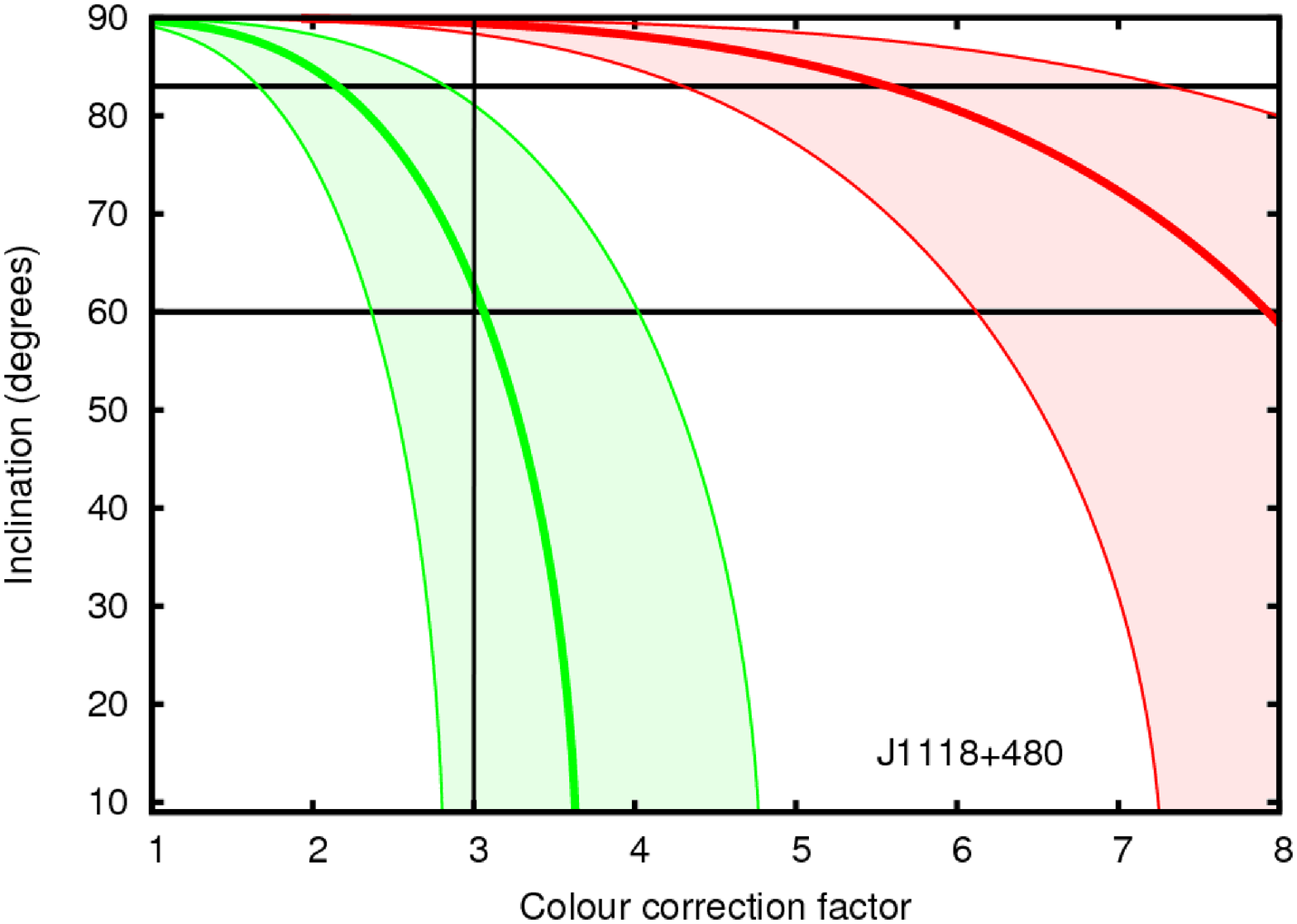}}
}
\rotatebox{0}{
{\includegraphics[width=225pt,clip ]{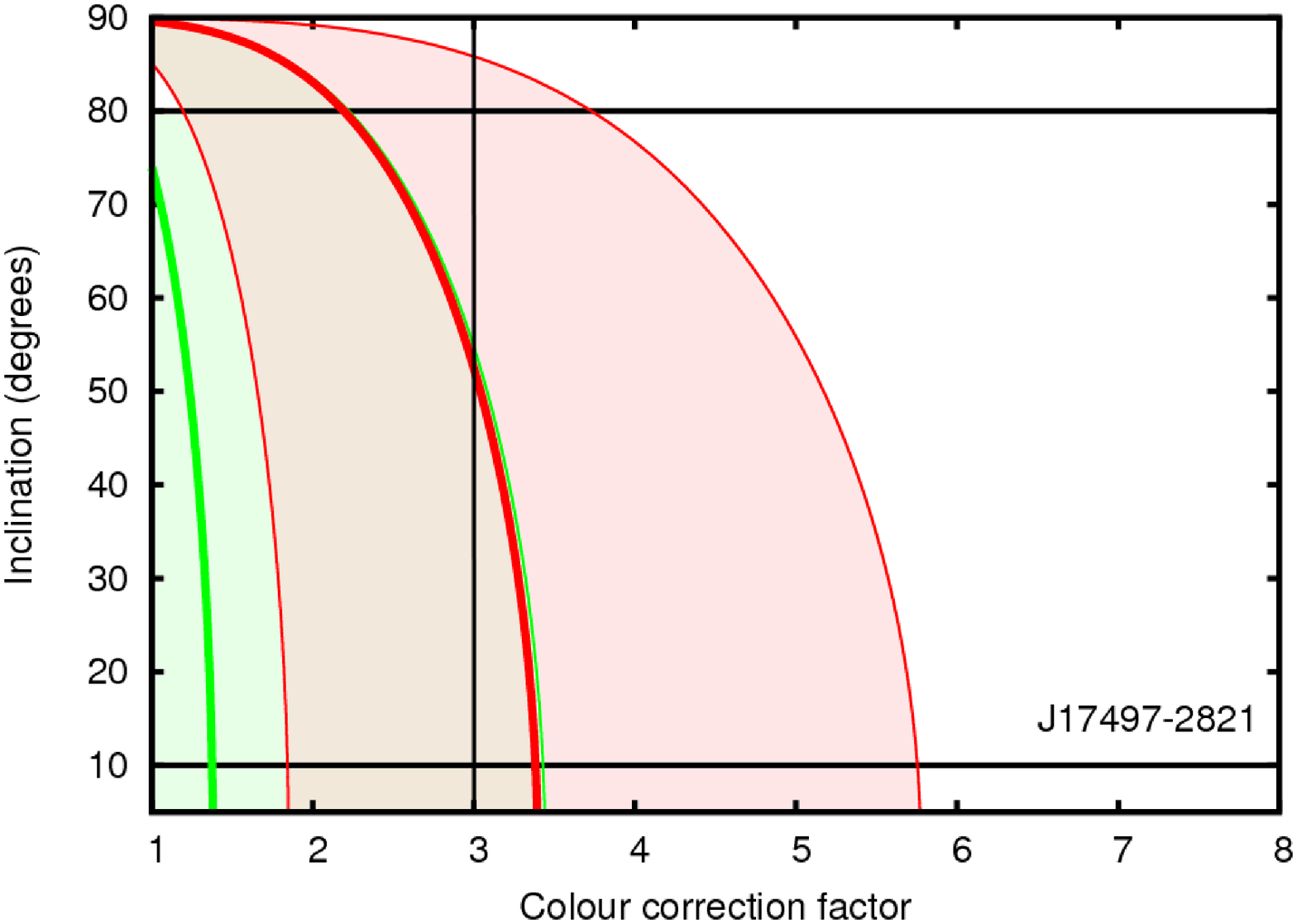}}
}
\rotatebox{0}{
{\includegraphics[width=225pt,clip ]{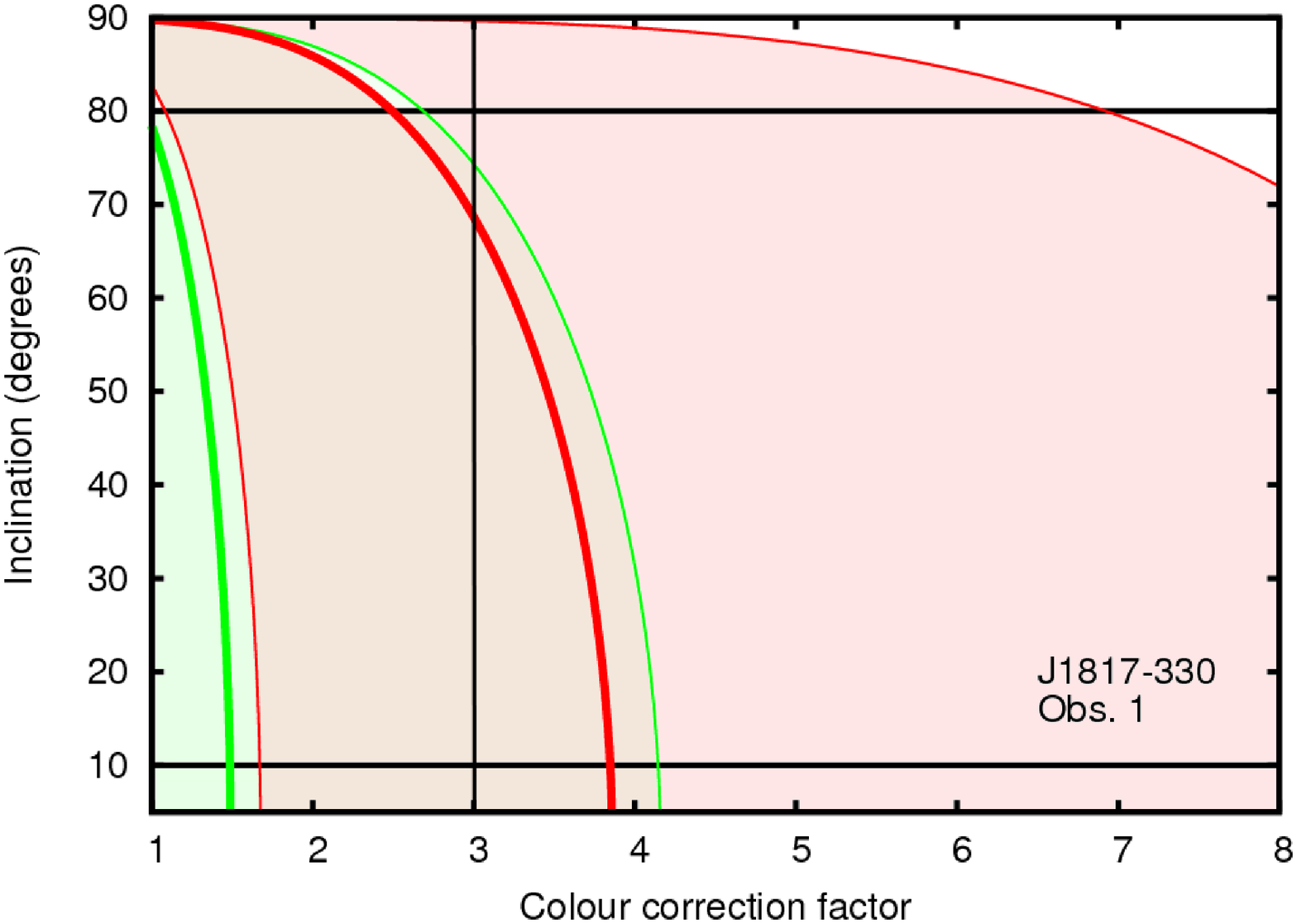}}
}

\caption{ Inclination vs colour correction factor $f$ for the various
  sources. The red and green curves are for a disk with inner radius
  of 100 and 6\rg respectively. The shaded regions shows the 3$\sigma$ error
  range. The upper and lower limits on the inclination are given by
  the solid horizontal lines and the maximum expected colour
  correction factor is show by the solid vertical line. For the model
  to be physically consistent the shaded region needs to be enclosed
  within the inclination range and to the left of the vertical
  line. This criteria rules out a disk at 100\rg\ for all sources
  other than J17497 and J1817. For simplicity only the first
  observations of Cyg{\thinspace X-1} and J1817 are shown.}
 \label{fig_inc_vs_beta}

\end{figure*}

\begin{table*}
  \caption{Results of fits with the \laor\ line profile and a powerlaw above 3\kev. GX 339-4 and J1650 were fit in the 3--12\kev\ range. For J1655 we used \rxte\ up to 20\kev\ and J1753 covered the full 3--100\kev\ range. For Cygnus X-1 only the XIS data were used up to 10\kev. The latter fit also included a narrow Gaussian line   at 6.4\kev. The upper energy limits were chosen so as to exclude the  prominent Compton hump present in most sources. The values of \nh\ are frozen at that shown in Table \ref{table_diskbb} and are given in units of $10^{21} \cm^{-2}$  . }

\begin{tabular}{lccccccccc}                
  \hline
  \hline
  Parameter & \nh\ & $\Gamma$ & $N_{PL}$ & $E_{Laor}$ (\kev) & $q$ & $\theta$ (degrees) &\rin\ (\rg)& $N_{Laor} (\times 10^{-3}$) & $\chi^2/\nu$\\
  J1650-500 &0.556&$2.17\pm0.01 $&$3.70^{+0.07}_{-0.01}$&$6.97_{-0.09}$&$7.8^{+0.5}_{-0.4}$&$65^{+5}_{-1}$&$1.45^{+0.03}_{-0.04}$&$37^{+3}_{-2}$&1352.2/1405 \\
  \gx &0.495&$1.671^{+0.012}_{-0.005} $&$0.225^{+0.004}_{-0.003}$&$6.93^{+0.04}_{-0.21}$&$3.2\pm0.5$&$10^{+17}$&$2.4^{+0.3}_{-0.5}$&$3.6^{+0.3}_{-0.8}$&1906.4/1786\\
  Cygnus X-1 (1)&0.53&$1.71\pm0.01 $&$1.75\pm0.02$&$6.97_{-0.04}$&$3.4^{+0.5}_{-0.3}$&$20^{+4}$&$6.3^{+1.4}_{-0.8}$&$6.6\pm1.0$&647.9/636 \\
  Cygnus X-1 (2)&0.50&$1.70\pm0.01 $&$1.48^{+0.02}_{-0.03}$&$6.97_{-0.08}$&$3.6^{+0.9}_{-0.6}$&$20^{+3}$&$7.1^{+2.3}_{-1.5}$&$5.9\pm1.0$&561.3/549  \\
  J1753.5-0127 &0.197&$1.60\pm0.01 $&$0.056\pm0.001$&6.4(f)&$2.2^{+0.4}_{-0.3}$&$67^{+8}_{-4}$&$2.4^{+1.8}_{-1.2}$&$0.14\pm0.05$&1450.8/1467 \\
  J1655-40 &0.63&$1.74^{+0.1}_{-0.2}$&$0.144^{+0.003}_{-0.005}$&$6.47^{+0.07}_{-0.06}$&$2.7\pm0.3$&$86.0\pm0.2$&$1.5^{+0.2}_{-0.1}$&$2.3^{+0.2}_{-0.4}$&921.1/922 \\
  \hline
  \hline
\end{tabular}
 
\end{table*}

\begin{table*}
  \caption{Results of fits to the full energy range using the
    reflection model \reflionx. Both the Compton hump and the iron line
    profile are modelled self-consistently. The inner radius is found
    from the blurring of the reflection component. The values of \nh\
    are frozen at that shown in Table \ref{table_diskbb} and are given
    in units of $10^{21} \cm^{-2}$ }
\begin{tabular}{lccccccccc}                
  \hline
  \hline
  Parameter & \nh\ & $\Gamma$ & $\xi\ ({\rm erg}\cm\s^{-1})$ & {\it q}  & \rin\ (\rg) & $\theta$ (degrees) &  $N_{PL}$& $N_{Reflionx}$ $(\times 10^{-6}$) & $\chi^2/\nu$\\
  J1650-500  &0.556&$2.19\pm0.02 $&$620^{+130}_{-45}$&$7.7^{+0.2}_{-0.3}$&$1.3\pm0.1$&$70_{-1}$&$2.5^{+0.2}_{-0.1}$&$102^{+11}_{-13}$&1441.5/1461\\
  GX 339-4 &0.495&$1.72^{+0.01}_{-0.02} $&$500^{+40}_{-90}$&$3.0\pm0.1$&$2.1\pm0.3$&$28^{+2}_{-18}$&$0.188^{+0.004}_{-0.006}$&$12^{+2}_{-3}$&1935.6/1839 \\
  J1655-40   &0.63&$1.71^{+0.01}_{-0.02} $&$240^{+17}_{-15}$&$2.72^{+0.7}_{-0.5}$&$1.38^{+0.6}_{-0.1}$&$87_{-1}$&$0.118^{+0.001}_{-0.003}$&$12^{+2}_{-1}$&957.7/966 \\
\hline
\hline
\end{tabular}
 
\end{table*}

\subsection{Can a disk at 100\rg\ radii explain the data?}
\label{innerrad_diskpn}

In the previous section we have shown using two well known multicolour
disk models that when the physical parameters of the systems are used
in conjunction with continuum modelling, the accretion disk likely
extends close to the ISCO in all sources investigated.

As previously mentioned, the ADAF model for accretion flow in the
low-hard state has two distinct zones, with the inner part being
modelled as a hot, optically thin advection dominated region while the
outer part consists of a standard Shakura-Sunyaev disk. The transition
radius is usually assumed to be $R_{tr}\sim${\it O}($10^3$)\rg\ (Esin
et al. 1997), however Wilms et al. (1999) and Esin et al. (2001) have
shown that for \gx\ and J1118 this radius is more likely to be
200--400\rg\ and $\approx$110\rg\ respectively. In the work that follows
we will systematically investigate the possibility of having a disk
truncated at 100\rg\ in the sample of systems in the low-hard state
presented here.

The model \diskpn\ (Gierlinski et al. 1999) is a further development
of the \diskbb\ model taking into account the torque-free inner
boundary condition{\footnote {It was in fact this work that pointed
    out the non-zero torque nature of the \diskbb\ model.}}. This
model has three parameters: The maximum colour temperature of the disk
($T_{col}$) in units of \kev\, the inner disk radius, \rin, and the
normalisation which is defined as $m^2cos\theta/(D^2f^4)$, where $D$ is
the distance to the source in \kpc\ and all other symbols are similar
to that of \diskbb\ and \ezdiskbb.  Given that the inner radius is now
a parameter of the model, we perform fits where this value is fixed at
both 6 and 100\rg.  As can be seen from Table 5, models with and
without disk truncation give equally satisfactory fits. The
temperature obtained in both cases are in remarkable agreement and
only the normalisation differ significantly. 

The normalisation of \diskpn, as well as of any other multicolour disk
model, conveys important physical information and can be used to set
apart different, often contradictory, interpretations. Using the
constraints on the masses and distances to the various sources (Table
1) together with the \diskpn\ normalisations for both \rin=6 and
100\rg\ we plot the inclination as a function of the colour correction
factor for each source (Fig.{\thinspace 8}). This factor is usually
assumed to be a constant close to 1.7 for a wide range in luminosity
(Shimura \& Takahara 1995). Davis et al. (2005)
showed that below a colour temperature of $\sim$1\kev\ the colour
correction factor is indeed a constant close to 1.7. Above this
temperature disk ionisation leads to a slight increase in $f$ however
it is found to be consistently below $\sim2.2$ (Davis, Done \& Blaes
2006; Done \& Davis 2008). Contrary to this, it was suggested by
Merloni, Fabian \& Ross (2000) that $f$ is a relatively strong function of accretion rates varying from $\approx$1.7--3.

The solid red and green curves in (Fig.{\thinspace 8}) shows the
dependence of the inclination on the colour correction factor for an
accretion disk at 100 and 6\rg\ respectively. The errors, shown by the
shaded regions, are the 3$\sigma$ errors estimated by Monte Carlo
simulations assuming a uniform distribution of the mass and distance
to the various sources (see Table 1) and the \diskpn\ normalisation
shown in table 5. It can be seen that for most sources a disk
truncated at 100\rg\ requires a high colour correction factor which
is, in most cases, inconsistent with the upper limit of three (solid
vertical line) set by Merloni, Fabian \& Ross (2000). We show in
\S\ref{discussion_diskbb} that this value is in fact likely to be
below $\approx2.4$ and the value of 3 used here is a conservative upper
limit. J17497 and J1817 are the only sources where a truncated
accretion disk is not ruled out due to the weak constraints on their
masses, distances and inclinations. However, we have seen from the
previous section that the data for these sources are suggestive of a
disk extending to within 10\rg.

\subsection{Inner radius from disk reflection}
\label{innerrad_reflection}

So far, we have only focused on the thermal component which in the
low-hard state is predominant below $\sim$2\kev. In the following
section we will divert our attention to the reflection features
present in five of the sources in our sample. To this end we will ignore the energies below 3\kev\ in all spectra
and, where possible, extend the high energy using either \rxte\ or
\suzaku\ PIN data. A detailed analyses of each individual source is
presented below.

\subsubsection{J1650-500}

 We have initially added the \rxte\ PCA data up the 12\kev\ to the \xmm\ spectrum in order to have a
clearer view of the continuum. The absorbing column density was fixed
at the value quoted on Table 3. An absorbed powerlaw does not provide
a satisfactory fit ($\chi^2/\nu=1988.1/1411$), with the bulk of the
residuals coming from the iron line region.

\begin{figure*}
\rotatebox{90}{
{\includegraphics[width=140pt, angle=180]{figure_j1650_xmm_pca_3_12kev_bestfit.ps}}
}
\hspace{0.5cm}
\rotatebox{90}{
{\includegraphics[width=140pt, clip]{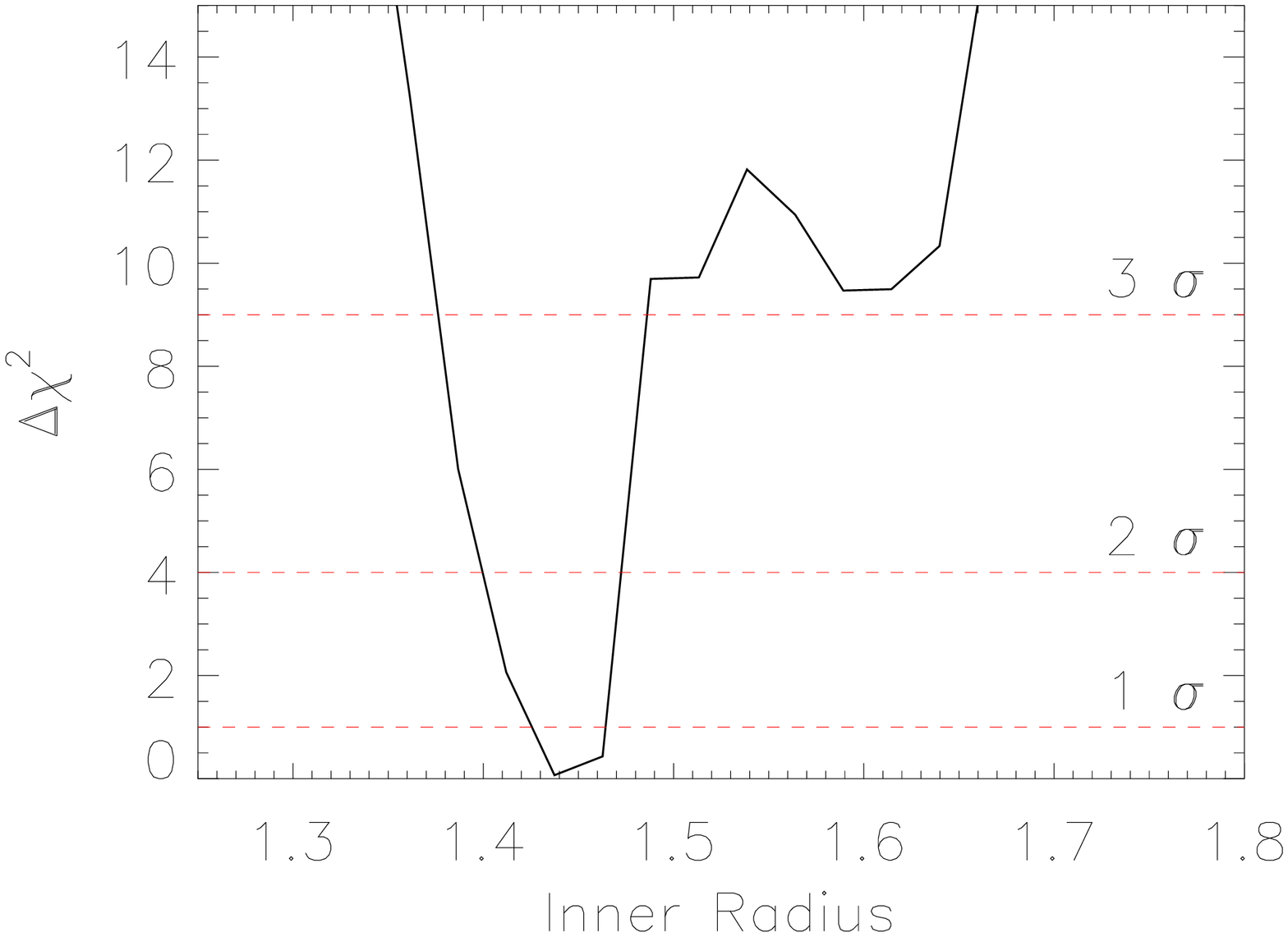}}
}
\rotatebox{270}{
{\includegraphics[width=140pt]{figure_j1650_xmm_pca_3_12kev_extended.ps}}
}
\begin{center}
  \caption{{\it Top-left:} Best fit model for J1650-500 in the
    3--12\kev\ energy range (Table 6). The \xmm\ data are shown in
    black with \rxtepca\ data shown in red.  {\it Top-right:}
    Goodness--of--fit variation as a function of inner radius. The
    horizontal lines mark the 1, 2, and 3$\sigma$ level. A disk
    radius greater than 6\rg\ is excluded at the 5$\sigma$ level (not
    shown). {\it Bottom:} Ratio of the best fit model above to the
    full, extended data. The HXTE (green) flux at 25\kev\ is
    normalised to match that of the PCA. The data have been re-binned
    for display purposes only.  }
\end{center}

\rotatebox{270}{
{\includegraphics[width=140pt]{figure_j1650_xmm_pca_3_100kev_reflion.ps}}
}
\hspace{1cm}
\rotatebox{270}{
{\includegraphics[width=140pt]{figure_j1650_xmm_rxte_0_100_bestfit_model_extended.ps}}
}
\caption{ {\it Left:} The best-fit reflection model for J1650-500. {\it Right:} Ratio of the best fit
  reflection model to the 0.6--100.0\kev\ range. The strong soft
  excess clearly demonstrates the requirement for a further
  thermal-disk component.}

\end{figure*}

\begin{figure*}
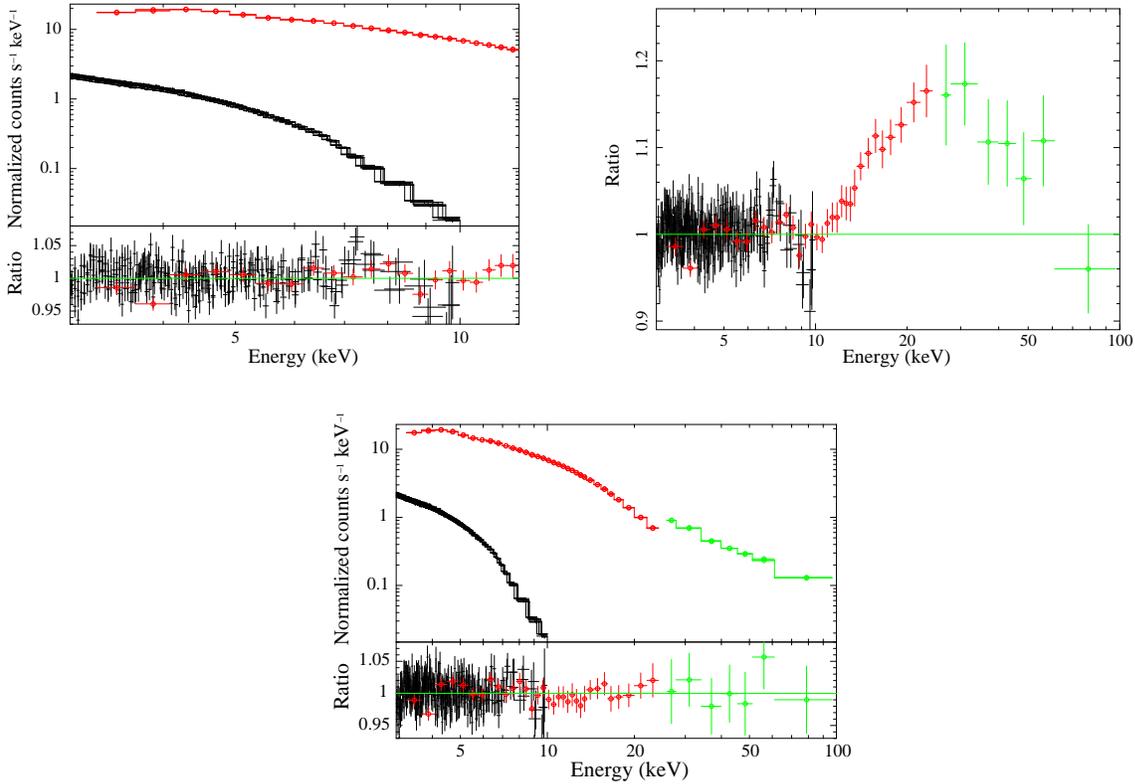

\begin{center}
\rotatebox{270}{
{\includegraphics[width=140pt]{figure_gx339_xmm_rxte_0_12_bestfit.ps}}
}
\hspace{0.5cm}
\rotatebox{270}{
{\includegraphics[width=140pt]{figure_gx339_xmm_rxte_0_12_bestfit_extended100kev.ps}}
}
\vspace{0.5cm}

\rotatebox{270}{
{\includegraphics[width=140pt]{figure_gx339_xmm_rxte_0_100_bestfit_reflion.ps}}
}

\caption{{\it Top-left:} Best fit model for \gx\ in the 3--12\kev\
  energy range (Table 6). The \xmm\ and \rxte-PCA data are shown in
  black and red respectively.  {\it Top-right:} Data/model ratio for
  the above model extended to 100\kev. The HXTE (green) flux at
  25\kev\ is normalised to match that of the PCA (red). {\it Bottom:}
  Best fit reflection model (Table 7). }
\end {center}
\end{figure*}

The standard and most generally used model for a relativistic line
profile around a rotating black hole is the \laor\ model in \xspec\
(Laor 1991). It describes a broad line from an accretion disc
surrounding a rotating Kerr black hole, with an emissivity profile
described by a power-law of the form $\epsilon_{r} = r^{-q}$. We use
this model to account for the broad residuals seen at the iron line
region. The \laor\ model is calculated with the presumption of a
maximally rotating black hole. Relativistic line models in which the
black hole spin is a free parameter are now available (KYRLINE and
KERRDISK models; Dovciak, Karas \& Yaqoob 2004 and Brenneman \&
Reynolds 2006 respectively) however we choose to use the well
established \laor\ model as we are not interested in the precise value
of the inner radius/spin and to allow for a more direct comparison
with previous work.  The outer disk radius in the \laor\ model was
fixed at the maximum allowed value of 400\rg\ and the inclination was
constrained to the values shown in Table 1 (47--70 degrees) for
J1650. The line energy was constrained to 6.4--6.97\kev. This
resulted in a marked improvement over the previous fit with
$\Delta\chi^2/\Delta\nu=-451.1/5$. A better fit is achieved when we
allow the photon indices of the \xmm\ and \rxte\ observations to
differ by $\le0.15$.  Miller (2009) has shown that small differences
in the value of $\Gamma$ between various X-ray instruments are
expected and we allow such deviations in all combined \rxte--\xmm\
analyses. The best fit ($\chi^2/\nu=1352.2/1405$) in the 3--12\kev\
range is shown in Fig. {\thinspace 9} (Top-left) and detailed in Table
6. Fig. {\thinspace 9} (Top-right) shows a plot $\Delta\chi^2$ versus
inner radius obtained with the command \steppar\ in \xspec. Based on this analyses, the accretion disk in J1650 is found to be within 6\rg\ at more than the 5$\sigma$ level of
confidence (Fig. 9).

Figure 9 (Bottom) shows the best fit mentioned above extended to
100\kev\ after the addition of HXTE data. The powerlaw indices of both
\rxte\ instruments are tied to each other. To make this figure we have
normalised the HXTE data so that the flux at 25\kev\ matches that of
the PCA. The large excess peaking at $\sim$30\kev\ is the ``Compton
hump'' associated with reflection of hard X-rays by a cool accretion
disk. In order to model the full spectra we will use the reflection
model \reflionx\ (Ross \& Fabian 2005) blurred with the same kernel
used in the \laor\ model (\kdblur\ in \xspec). The best \reflionx\ fit
covering the full energy band is shown in Fig. 10 (Left) and
detailed in Table{\thinspace 7}. When using the reflection model \reflionx\ the Fe-\ka\
line, as well as the Compton hump are modelled self-consistently. It
should be noted that the refection component cannot account for the
soft-excess explored in the previous chapters in any of the sources
investigated here. To emphasise this point we show in Fig. 10 (Right)
the data/model ratio of the best fit reflection model of J1650
extrapolated to the soft energy range. It is clear  from this
figure that a soft disk component is still required.

\begin{figure}
\begin{center}
\rotatebox{90}{
{\includegraphics[width=160pt, clip]{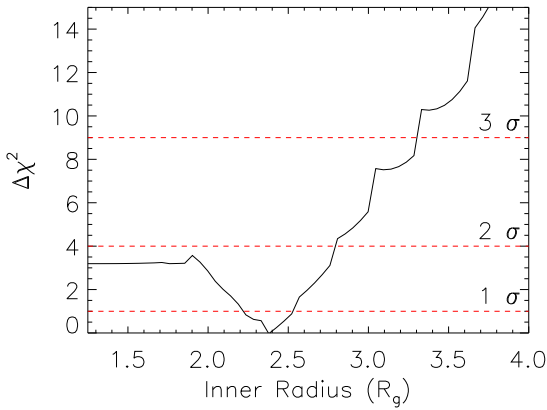}}
}

\caption{ Goodness--of--fit variation as a function of inner radius
  for \gx. The horizontal lines mark the 1, 2, and 3$\sigma$
  level. A disk radius greater than 6\rg\ is excluded at the 5$\sigma$
  level (not shown). }
\end {center}
\end{figure}

\subsubsection{GX339-4 }

\begin{figure*}
\begin{center}
\rotatebox{270}{
{\includegraphics[width=140pt]{figure_cygx1_3_10_laor.ps}}
}
\hspace{1cm}
\rotatebox{270}{
{\includegraphics[width=140pt]{figure_cygx1_3_10_laor_ga.ps}}
}
\vspace{0.2cm}
\caption{ {\it Left:} Data/model ratio for Cyg{\thinspace X-1} with a powerlaw and
  relativistic line. {\it Right} With the addition of a narrow
  Gaussian at 6.4\kev. Obs 1 and 2 are shown in black and red
  respectively. }

\rotatebox{270}{
{\includegraphics[width=140pt]{figure_cygx1_contour_rin_emmi_obs5.ps}}
}
\hspace{1cm}
\rotatebox{270}{
{\includegraphics[width=140pt]{figure_cygx1_contour_rin_emmi_obs6.ps}}
}
\vspace{0.2cm}
\caption{Emissivity versus inner radius contour plot for Cyg{\thinspace X-1}. The
  68, 90 and 95 per cent confidence range for two parameters of
  interest are shown in black, red and green respectively. It can be
  seen that for the full range of the emissivity the inner radius is
  constrained between approximately 5--8\rg\ (Obs 1; Left) and 5--9\rg\
  (Obs 2; Right) at the 90 per cent confidence level for two
  parameters . }

\rotatebox{90}{
{\includegraphics[width=140pt]{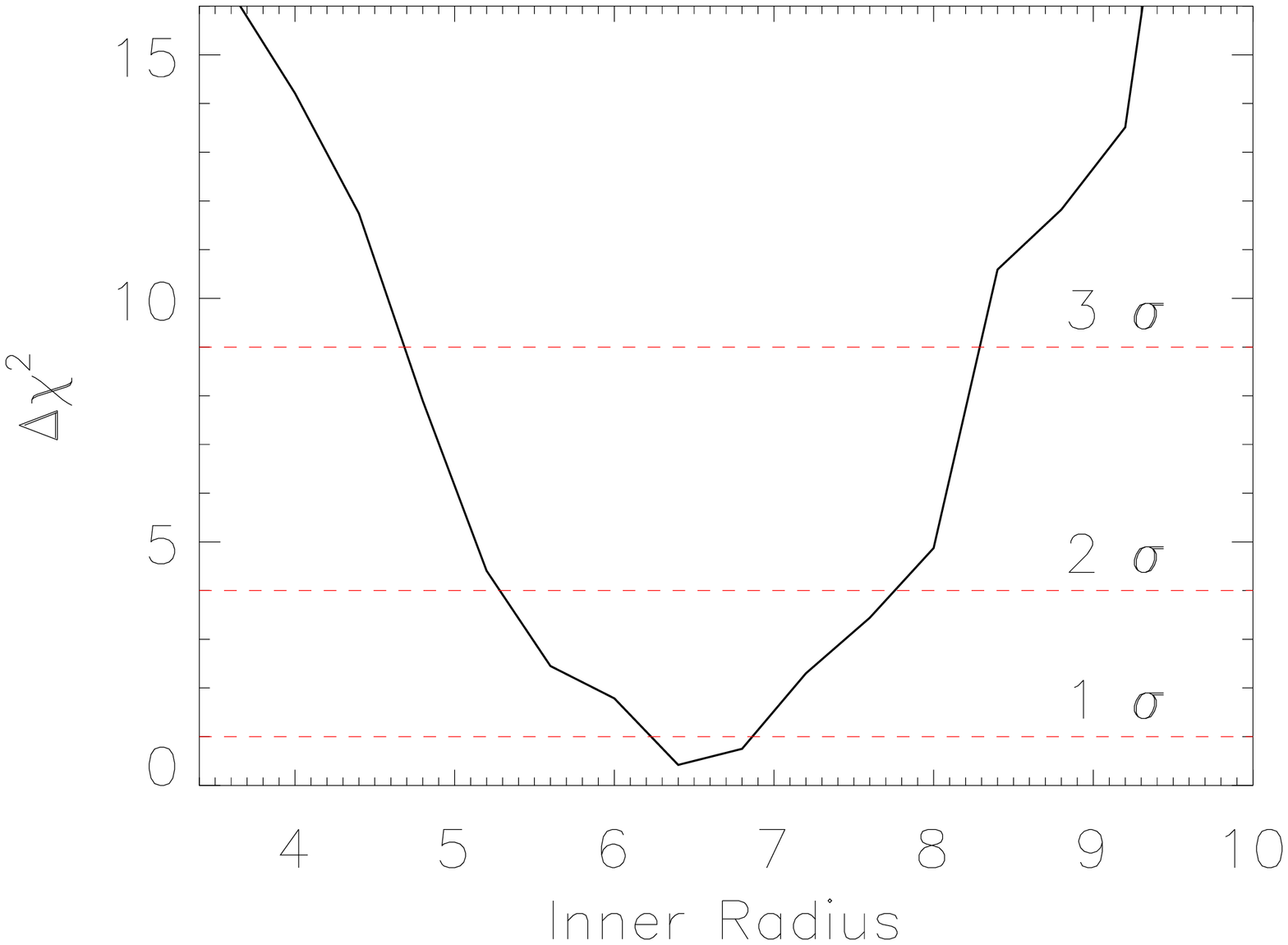}}
}
\hspace{1cm}
\rotatebox{90}{
{\includegraphics[width=140pt]{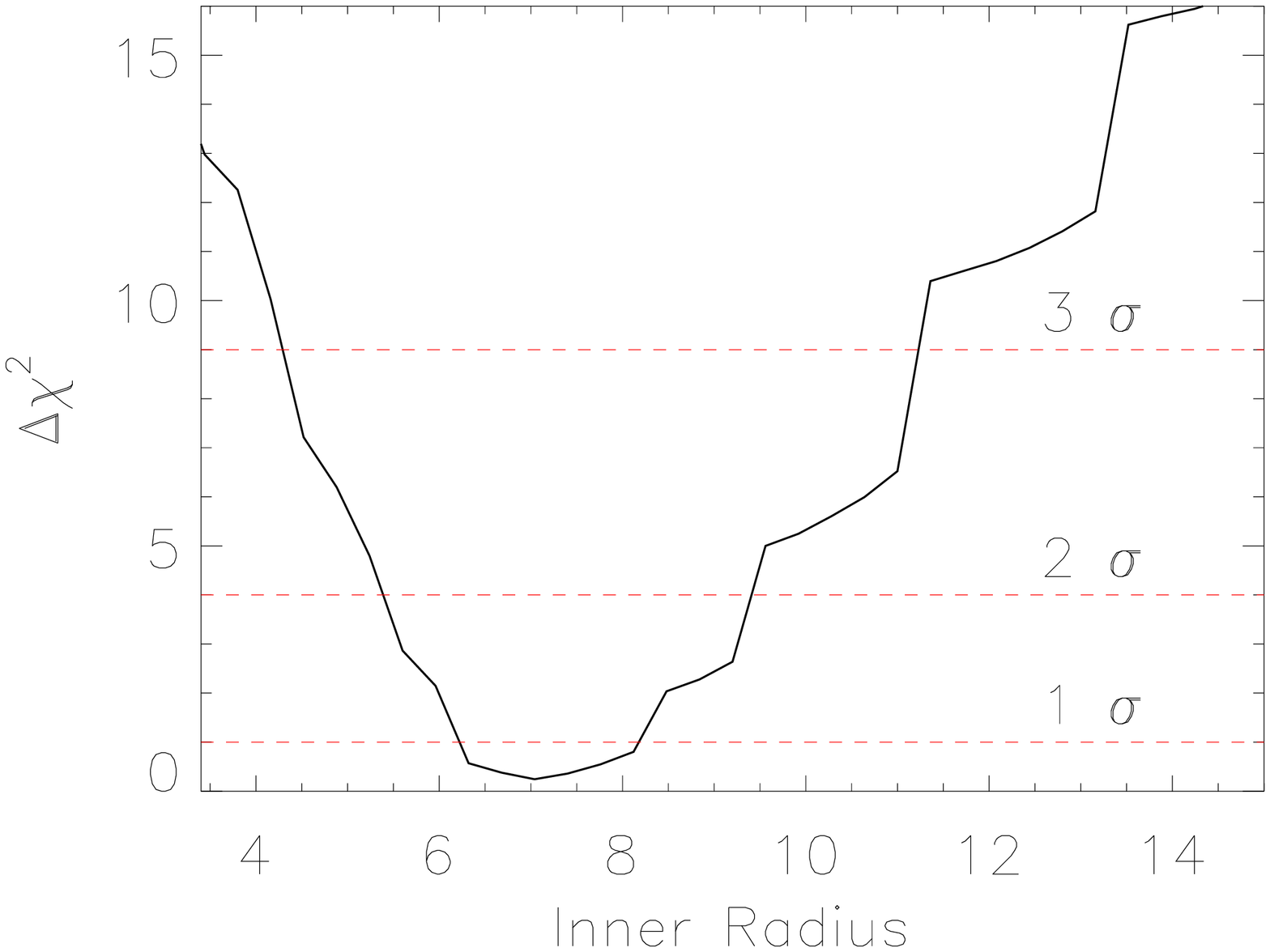}}
}
\caption{ {\it Left:} Goodness--of--fit variation as a function of
  inner radius for Cygnus X-1 (Obs 1). {\it Right} Same but for
  Obs. 2.  The horizontal lines mark the 1, 2, and 3$\sigma$
  level. }
\end{center}
\end{figure*}

The presence of strong iron line emission in the \xmm\ spectrum of
\gx\ was previously seen in Fig. 1.  A powerlaw fit over the
3--12\kev\ range results in a poor fit with
$\chi^2/\nu=2617.7/1791$. This is dramatically improved by the
addition of the \laor\ line model resulting in
$\Delta\chi^2/\Delta\nu=-711.3/5$ with an inner radius approaching
that expected for a maximally rotating black hole (Fig. 12).

Figure 11 (Top-left) shows the best fit model over the 3--12\kev\ energy
range. The fit extended to the full 3--100\kev\ range is shown in
Fig. 11 (Top-right) with the HXTE flux normalised to match that of
the PCA at 25\kev. Similarly to the spectra of J1650 this source shows
the presence of a strong Compton reflection hump as expected from
a source with strong fluorescence emission. Replacing the \laor\
component with the blurred-reflection model results in an acceptable
fit to the full energy range (Fig. 11 Bottom) with the inner
radius again implying a maximally rotating black hole. Based on the
extent of the gravitational blurring of the reflection features in
\gx, an accretion disk with an inner edge at a distance greater
than 6\rg\ is excluded at more than the 5$\sigma$ confidence level.

\begin{figure*}
\rotatebox{270}{
{\includegraphics[width=140pt]{figure_j1655_xis_pin_3_20_laor_bestfit_fi.ps}}
}
\rotatebox{270}{
{\includegraphics[width=140pt]{figure_1753xmm_rxte_bestfit.ps}}
}
\caption{ Best--fit model consisting of powerlaw plus relativistic
  line. {\it Left:} J1655-40 in the 3--20\kev\ range. {\it Right:}
  J1753.5-0127 in the 3--100\kev\ range. The HXTE data is omitted for
  display only. }

\rotatebox{270}{
{\includegraphics[width=140pt]{figure_j1655_xis_pin_3_20_laor_extended70kev_fi.ps}}
}
\rotatebox{270}{
{\includegraphics[width=140pt]{figure_j1655_xis_pin_3_70_reflion.ps}}
}
\caption{ {\it Left:} Data/model ratio for J1655-40 with a
  model consisting of powerlaw plus relativistic line fitted in the
  3--20\kev\ range. {\it Right:} Best--fit reflection model in the full
  3--70\kev\ range }

\rotatebox{90}{
{\includegraphics[width=150pt]{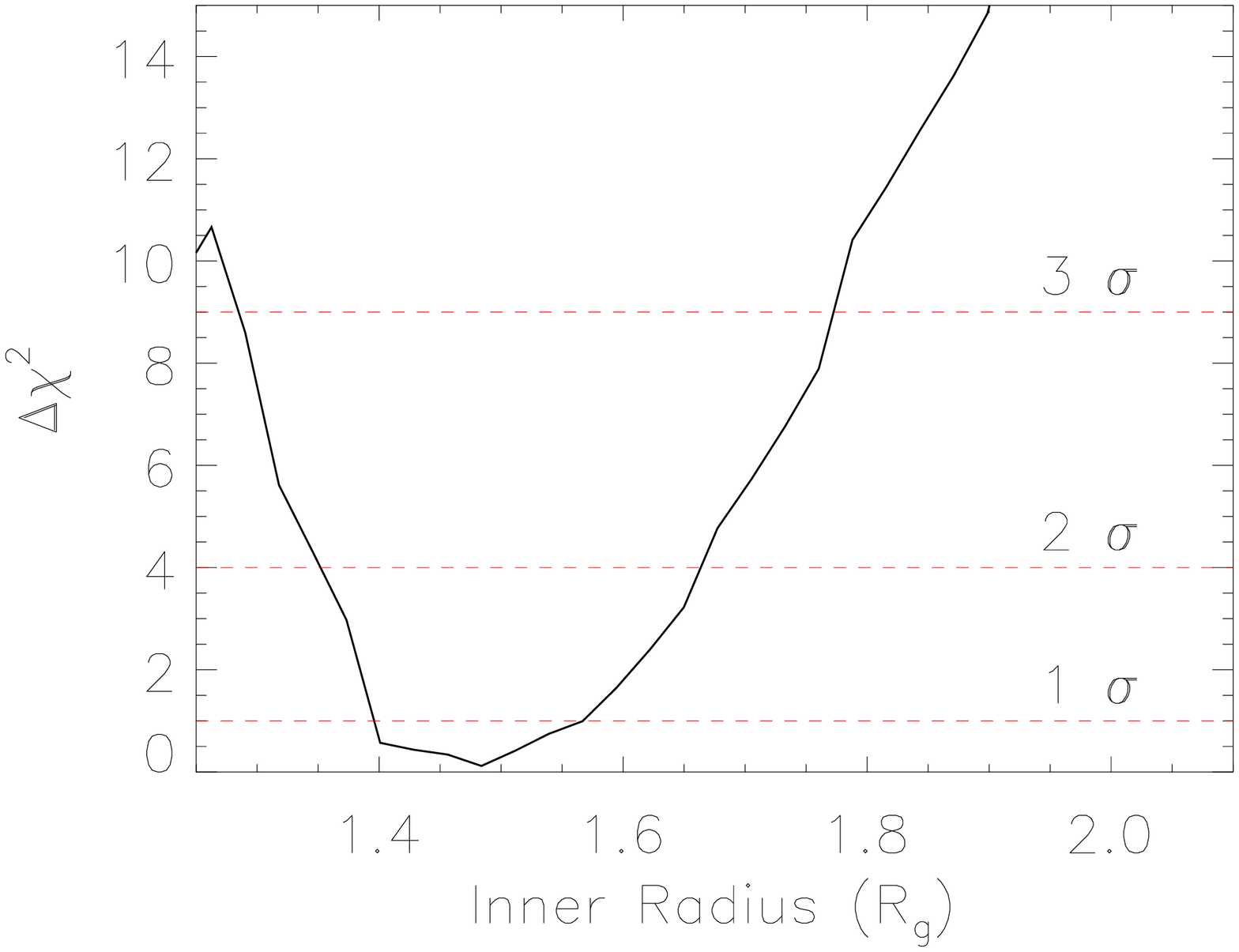}}
}
\rotatebox{90}{
{\includegraphics[width=150pt]{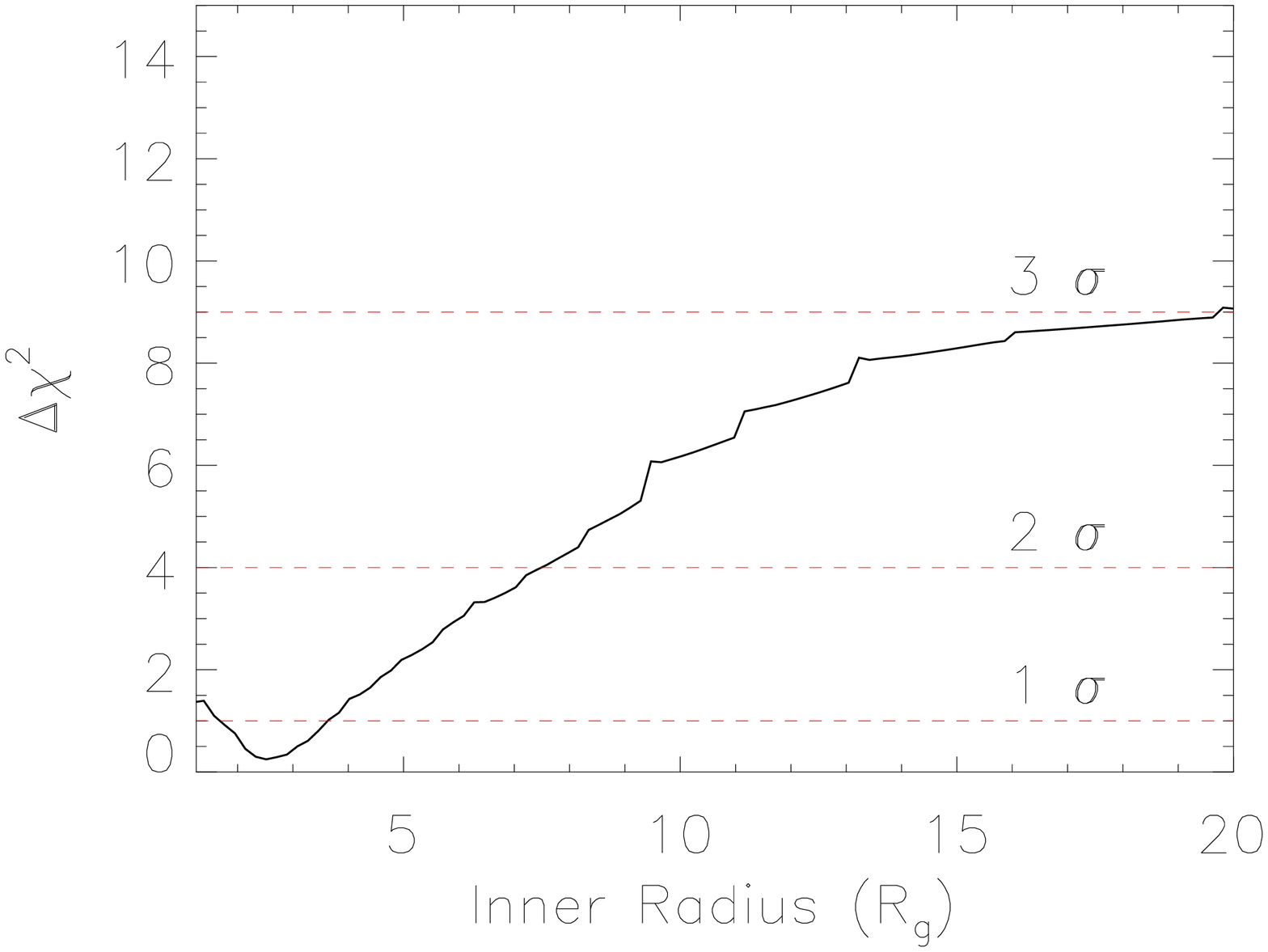}}
}

\caption{{\it Left} Goodness--of--fit variation as a function of inner
  radius for J1655-40. {\it Right} Similar to the left but for J1753.5-0127.  }

\end{figure*}

The value for the inner radius found using the full spectra
($2.0\pm0.3$\rg) is similar to that obtained in the narrower
(3--12\kev) range from the iron-line shape alone
($2.4^{+0.4}_{-0.5}$\rg; Fig 12). This emphasises the robustness of
spin measurements made from the shape of such strong
reflection features. The result presented for the innermost extent of
the accretion disk in the low-hard state is similar to that found for
\gx\ in both the intermediate state (2.0--2.2\rg; Miller et al. 2008)
and the joint fitting of the low-hard and very-high state
(2.0--2.1\rg; Reis et al. 2008) which confirms that the disk remains
stable during these three spectral states.

\subsubsection{Cygnus X-1}

The profile around the iron-\ka\ region in Cyg{\thinspace X-1} cannot be modelled with a single relativistic line (Fig 13; Left). However,
satisfactory fits are obtained for both observations by the addition
of a narrow Gaussian at 6.4\kev\ together with the \laor\ line. This
resulted in $\chi^2/\nu=647.9/636$ and $\chi^2/\nu=561.3/552$ for
observations 1 and 2 respectively. The Gaussian is required at more
than the 5$\sigma$ level of confidence (F-value = 28.0 and Prob. =
1.7$\times10^{-7}$).  In both observations the inner radii and the
emissivity indices are well constrained, and it can be seen from
Fig. 14 that the subtle correlation between these parameters does not
provide a satisfactory solution with large radii for any physical
values of emissivity.  Table 6 details the various parameters in the
model. The inner radii are constrained to $6.4^{+1.4}_{-1.0}$ and
$7.1^{+2.3}_{-1.5}$\rg\ (90\% confidence) in Obs. 1 and 2
respectively, as shown in Fig. 15.

\subsubsection{J1655-40 and J1753.5-0127}

Figure 16 shows the  best fit model consisting of a powerlaw and
the \laor\ line profile for both J1655 (left) in the 3--20\kev\ energy
range and J1753 (right) in the full range. The relatively weak
reflection features in J1753 does not allow for a strong
constraint on the inner radius using simple phenomenological models,
with a disk at greater than 20\rg\ only being ruled out at the
3$\sigma$ level (Fig. 18; right). However, Reis et al. (2009b) showed
using a self-consistent reflection model that when the full energy
range (0.5--100.0\kev) is considered, the value of the inner radius is
constrained at approximately 2.5--5.0\      rg\ in agreement with the value
found in section 3.1 (Table 3). Extending the fit in J1655 to the full
3.0--70.0\kev\ range again shows the presence of a Compton hump
(Fig. 17; left) which is successfully modelled with the reflection
model \reflionx\ (Fig. 17; right). The inner accretion disk radius in
J1655 is constrained to less than 6\rg\ at more than 5$\sigma$
confidence (Fig. 18; left). Table 7 details the parameters found for
the fit over the full energy range using \reflionx.

In all sources investigated in this section the equivalent widths of
the Fe-\ka\ line as modelled with the \laor\ line profile have been
found to be consistently above 70\ev\ at the 90 per cent confidence
level. The only exception to this was J1753 where the
equivalent width is found to be approximately 50\ev. In the following
chapter we consolidate our results and discuss the implication they
have on models for accretion flow.

\section{Summary and Discussion}
\label{discussion}

In this work we have systematically analysed a sample of eight high
quality, stellar mass black hole spectra obtained with various CCD and
grating instruments. We show observational evidence for the
possible presence of an optically-thick accretion disk extending close
to the ISCO in all eight sources in the low-hard state. In half of the sources the evidence comes from {\it both} the highly significant
thermal component as well as blurred reflection features.

\subsection{Constraints from the thermal disk}
\label{discussion_diskbb}

In all the objects studied the presence of a low-energy soft excess is
seen when the spectra are fit with an absorbed powerlaw
(Fig.{\thinspace \ref{fig_ratio_topl}}). This excess is traditionally
modelled assuming thermal emission from an accretion disk where the
flux, and hence normalisation, is related to the disk inner radius
according to $N\propto$ \rin$^2$. In \S\S \ref{innerrad_diskbb} and
\ref{innerrad_diskpn} we have used this information along with
constraints on the inclinations, masses and distances to the sources
found in the literature to obtain the inner radius of accretion  for the various sources. The results are shown in Figs.{\thinspace
  \ref{fig_rin_lumin}--\ref{fig_rin_kt}}.  For six out of the eight
sources investigated the 3$\sigma$ upper limit on their inner radii
are found to be below $\approx$10\rg\ and in all cases they are
consistent with extending to the inner-most stable circular orbit down
to luminosities of $\sim 5\times 10^{-4}L_{edd}$. The lack of
constraints in the physical parameters (mass, distance and
inclination) of J17497 and J1817 results in a large
uncertainty in the derived inner radii, however even when extreme
values are used for these parameters, the results are still consistent
with lack of disk truncation further than $\sim$100\rg\ for these
sources. All results presented above are robust to the choice of
multi-colour disk component used. We emphasise here that throughout this
work we have used the largest range in the physical parameters for the
various sources obtained from the literature. Even with this
conservative choice we have shown that the 3$\sigma$ upper limit on
the disk radius, as obtained from the flux of the thermal component,
does not require large disk truncation.

It is interesting to note that in all sources presented here the disk
temperature are found to be below $\approx$0.3\kev, with the majority
clustered at $\approx$0.2\kev. The disk flux found at these low accretion
rates results in an insignificant amount of flux visible to \rxtepca,
with its effective low energy cut-off of $\approx$3\kev. Zdziarski et al.
(2004) investigated the state transition of \gx\ using \rxte. Their
Figure 10 shows a sharp change in the inner radius - obtained via the
disk flux methodology - during the change to the low-hard
state. Similar conclusions based on the apparent change in the
normalisation of the \diskbb\ model have been presented by Belloni et
al. (1997) for the black hole binary GRS J1915+105 again using
\rxtepca\ data above 2.5\kev.

Using a further multi-colour disk model where the inner radius is a
free parameter we investigated in \S{\thinspace \ref{innerrad_diskpn}}
the possibility that the thermal component originates further than
100\rg\ as expected from ADAF models. Figure \ref{fig_inc_vs_beta}
shows that this conviction was excluded for all sources other than
J17497 and J1817. Our conclusions depends on the maximum
plausible value for the spectral hardening factor $f$. Hardening arise
at higher photon energies where the opacity of the disk atmosphere is
lower and photons emerge from deeper layers where gas is
hotter. Approximately, $f=T/T_{\rm eff}\simeq\tau^{1/4}$, where $\tau$
is the Thomson depth of the layer from which the photons emerge (Ross,
Fabian \& Mineshige 1992; Davis, Done \& Blaes 2006 in the high $\tau$
limit). The photons then need to scatter out through the outer layers
which are cooler than where they formed. This introduces a Compton
down-scattering break in the spectrum at an energy $\epsilon\sim
m_{\rm e} c^2/\tau^2$. There are therefore two competing effects:
seeing to deeper layers making the spectrum harder and Compton
down-scattering by intervening material making it softer. The maximum
value of $f$ occurs before the effects balance, which is at $f=(m_{\rm
  e} c^2/\epsilon)^{1/8}$. Therefore if we take $\epsilon\sim
0.5$\kev\ to mark the harder part of the thermal emission, $f<2.4$. An
absolute limit will depend on exactly which part of the spectrum is
being fitted. We consider that $f=3$ is a conservative upper limit.

The high significance of the thermal component in the various sources
studied here, coupled with the total flux and temperature range ($\approx
0.15-0.3$\kev) obtained from multi-colour disk models such as
\diskbb, strongly argue for the presence of a standard optically-thick
accretion disk extending close to the ISCO.

\subsection{Constraints  from reflection signatures}

Further evidence for an optically-thick disk extending to the ISCO in
the low-hard state comes from the presence of a broad, skewed Fe-\ka\
line and strong reflection hump observed in five out of the eight
sources presented. These reflection features arise due to
reprocessing of hard X-ray by the cooler accretion disk (Ross \& Fabian
1993). In the region surrounding a black hole the strong gravitational
potential causes these reflection features to become highly distorted,
with the degree of distortion depending on how far from the black hole
the emitting region is located. Therefore, the shape of the prominent
iron-\ka\ line can give a direct indication of the radius of the
reflection material from the black hole (Fabian et al. 1989, 2000;
Laor 1991). 

Section \ref{innerrad_reflection} shows the reflection signatures of
five sources. It is clear from Figs 9 to 18 that in all cases investigated the inner radius of emission, as obtained from the degree of gravitational blurring of these features, is consistent with the ISCO down to luminosities of $\sim 1.5\times 10^{-3}L_{edd}$. For \gx, J1650, J1655 and J1753 the results further imply that the central source is a rotating, Kerr black hole with the disk extending to less than 6\rg\ at 5$\sigma$
confidence in most cases. This is in agreement with results found from
prior studies in various  states (\gx: Reis et al. 2008;
Miller et al. 2008. J1650: Miller et al. 2002; Miniutti et
al. 2004. J1655: Diaz-Trigo et al. 2007; Reis et
al. 2009b. J1753: Miller et al. 2006; Reis et al. 2009b) which
supports the idea that the emitting area is not changing between these
states.  A non-rotating, Schwarzschild black hole is not ruled out for
Cyg{\thinspace X-1}, however a disk with an edge at greater than 20\rg\ is
excluded at over 5$\sigma$. 

For four out of the five sources where an iron-\ka\ emission line is seen, the equivalent width, as measured from the \laor\ line component, is found to be greater than 70\ev\  at the 90 per cent level of confidence. It was shown by George \& Fabian (1991) that for a neutral, optically thick  accretion disk extending to the ISCO the reflection fraction $R\sim W_{K\alpha}/180\ev$. Beloborodov (1999) later showed that reflection fractions as low as $\approx$0.3 can be achieved by the mild relativistic bulk motion of a corona away from an accretion disk.  Based on these arguments it can be seen that $W_{K\alpha}\gtrsim60\ev$ is still consistent with originating in an accretion disk extending to the ISCO.

\subsection{Radio jets and disk truncation}

The sample investigated here covers a large range of luminosities from
$\approx$0.05 to 1.5\% of the Eddington limit, however this is still an
order of magnitude larger than the predicted quiescent transition
value of $\approx$4$\times10^{-5}L_{Edd}$ (Gallo et al. 2003) below which the
black holes are thought to be jet-dominated. The presence of
radio-jets is usually associated with systems in the low-hard state
(see e.g. Fender 2001, 2006), and it has at times been  attributed to a
receding (ejected?) inner disk (see e.g. Belloni et al. 1997). In this scenario the inner part of the accretion flow in the low-hard state is advection dominated. The radio jet is then
quenched when the inner region is filled by an accretion disk, as is
the case in the high-soft state. Our findings suggests that jet production is not initiated at the point where the accretion disk starts to recede. Another possibility for the production of jets could be due to a change in the ratio of the energy dissipated in a corona to that
dissipated in the disk between the various states, an intrinsic change
in the mass accretion rate through the disk or even changes in the
vertical scale-height close to the central black hole.

It is still possible that below the luminosities studied here, an
advection flow might be present at which point the disk could be
truncated. This ADAF solution is usually agreed to be the dominant
emission process in the Galactic Center source Sgr A$^*$ (Narayan, Yi
\& Mahadevan 1995; Narayan et al. 1998) as well as various
low-luminosity AGNs (LLAGNs; Di Matteo et al. 2000, 2003). However, an
alternative, jet-dominated accretion flow explanation has also been
successfully applied to these sources (Sgr A$^*$: Falcke \& Biermann
1999; Falcke \& Markoff 2000; LLAGNs: Falcke et al. 2000). In order to
test whether the accretion disk recedes at luminosities below those
observed here, as well as the profound connection between accretion
disk and jets, we strongly encourage deeper observations of black
holes in the low-hard and quiescent states from instruments such as
\xmm, \suzaku, and in the future \ixo.

\section{Conclusion}

In this paper we have investigated a sample of stellar mass black holes in the canonical low-hard state. By systematically analysing their X-ray spectrum we have found that in all cases the accretion disk is consistent with being at the inner-most stable circular orbit down to luminosities as low as $\approx5\times10^{-4}L_{Edd}$. The main points and implication of this paper are summarised as follows.

\begin{enumerate}
 \item In all sources investigated the presence of an accretion disk is required at the 5$\sigma$ level of confidence. The temperature and flux  of this thermal component is consistent with the $L\propto T^4$ relation and with a geometrically thin, optically thick accretion disk extending to the ISCO.
 \item The presence of reflection features and  predominantly an  iron-\ka\ emission line with an equivalent width greater than $\sim$70\ev\ is detected in half of the sample. In all these cases, the broadness of the reflection features further suggest that the accretion disk is not highly truncated.
 \item Our findings suggest that transition to the low-hard state are driven by changes in the corona (perhaps related to jet formation) and not changes in the accretion disk, and
 \item Jet production is not initiated at the point where the disk recedes.
 \item Furthermore, we suggest the following strong and weak observation criteria for disk truncation:
\begin{itemize}
 \item The data must be able to rule-out both a broad iron-\ka\ line with an equivalent width  $\gtrsim$60\ev\ AND an effective disk temperature consistent with $L\propto T^4$ or
 \item The data must be able to rule-out either a broad iron-\ka\ line with an equivalent width  $\gtrsim$60\ev\ OR an effective disk temperature consistent with $L\propto T^4$.
\end{itemize}
\end{enumerate}

 Whilst the number of sources presented
here is small, there is a general trend that when the accretion disk
is statistically required the data suggests that the disk in the
low-hard state can remain at the ISCO. This result is contrary to the lore that the accretion disk in
the low-hard state is truncated at hundreds of gravitational radii, as
required in the strong-advection dominated interpretation.

\section{Acknowledgements}
We would like to thank Mike Nowak, Joern Wilms and Katja Pottschmidt
for the use of Cygnus X-1 data. We would further like to thank Katja
Pottschmidt for scheduling the Suzaku observations of the same source.
RCR acknowledges helpful comments from Ed Cackett, Abdu Zoghbi and
Greg Salvesen which greatly improved this work, as well as STFC for
financial support. ACF thanks the Royal Society.


\begin{thebibliography}{}

\bibitem{} Anders E. \& Grevesse, N., 1989, GeCoA, 53. 197A 

\bibitem{} Arnaud K.A., 1996, ASPC, 101, 17A

\bibitem{} Bardeen J.M., Press W.H. \& Teukolsky S.A., 1972, ApJ, 178,
  347

\bibitem{} Barrio F. E., Done C., Nayakshin S., 2003, MNRAS, 342, 557B

\bibitem{} Beloborodov A. M., 1999, ApJ, 510, L123

\bibitem{} Belloni T., Mendez M., King A. R., van der Klis M., van
  Paradijs J., 1997, ApJ, 479L, 145B

\bibitem{} Brenneman L.W., Reynolds C.S., 2006, ApJ, 652, 1028B

\bibitem{} Brocksopp C., McGowan K. E., Krimm H., Godet O., Roming P.,
  Mason K. O., Gehrels N., Still M., Page K., Moretti A., Shrader
  C. R., Campana S., Kennea, J., 2006, MNRAS, 365, 1203B

\bibitem{} Caballero-Garcia M., Fabian A. C., 2009, MNRAS submitted

\bibitem{} Cackett E. M., Miller J. M., Bhattacharyya S., Grindlay J. E., Homan J., van der Klis M., Miller M. C., Strohmayer T. E., Wijnands R., 2008, ApJ, 674, 15C


\bibitem{} Cadolle Bel, M., Rib{\'o}, M., Rodriguez, J., Chaty, S.,
Corbel, S., Goldwurm, A., Frontera, F., Farinelli, R., D'Avanzo, P.,
Tarana, A., Ubertini, P., Laurent, P., Goldoni, P., Mirabel, I.~F.,
2007, ApJ, 659, 549C



\bibitem{} Davis S. W., Blaes O. M., Hubeny I., Turner N. J. 2005,
  ApJ, 621,372D

\bibitem{} Davis S. W., Done C., \& Blaes O. M.,  2006, ApJ, 647, 525

\bibitem{} D{\'{\i}}az Trigo, M., Parmar A.~N., Miller J. M., Kuulkers  E., and {Caballero-Garc{\'{\i}}a} M.~D., 2007, AIPC, 924, 877


\bibitem{} Di Matteo T., Celotti A., Fabian A. C., 1999, MNRAS, 304, 809 

\bibitem{} Di Matteo T., Quataert E., Allen S. W., Narayan R., Fabian A. C,  2000, MNRAS, 311, 507D

\bibitem{} Di Matteo T., Allen S. W., Fabian A. C., Wilson A. S., Young A. J., 2003, ApJ, 582, 133D

\bibitem{} Done C., Gierlinski M., 2006, MNRAS, 367, 659

\bibitem{} Done C., Davis S. W, 2008, ApJ, 683, 389D


\bibitem{} Dotani T., Inoue H., Mitsuda K., Nagase F., Negoro H., Ueda
  Y., Makishima K., Kubota A., Ebisawa K., Tanaka Y., 1997, ApJ, 485L,
  87D

\bibitem{} Dovciak M., Karas V., \& Yaqoob T., 2004, ApJ, 153, 205

\bibitem{} Esin A. A., McClintock J. E., Narayan R., 1997, ApJ, 489, 865

	
\bibitem{} Esin, A. ., McClintock J. E., Drake J. J., Garcia M. R.,
  Haswell C. A., Hynes R. I., Muno M. P., 2001, ApJ, 555, 483

\bibitem{} Fabian A.C., Rees M.J., Stella L., White N.E., 1989, MNRAS,
  238, 729

\bibitem{} Fabian A.C., Iwasawa K., Reynolds C.S., Young A.J., 2000,
  PASP, 112, 1145

\bibitem{} Fabian A. C., Zoghbi A., Ross R. R., Uttley P., Gallo L. C., Brandt W. N., Blustin A. J., Boller T., Caballero-Garcia M. D., Larsson J., Miller J. M., Miniutti G., Ponti G., Reis R. C., Reynolds C. S., Tanaka Y., Young, A. J., 2009, Nature, 459, 540F

\bibitem{} Falcke H., Biermann P. L., 1999, A\&A, 342, 49F

\bibitem{} Falcke H., Markoff S., 2000, A\&A, 362, 113F

	
\bibitem{} Falcke H., Nagar N. M., Wilson A. S., Ulvestad J. S., 2000, ApJ, 542 197F

\bibitem{} Fender R. P., 2001, MNRAS, 322, 31F


\bibitem{} Fender R. P., Belloni T. M., Gallo, E., 2004, MNRAS, 355, 1105F

\bibitem{} Fender R. P., 2006, in Lewin W. H. G., van der Klis M., eds, Compact Stellar X-ray Sources.Cambridge Univ. Press, Cambridge, p. 381

\bibitem{} Frontera F., Zdziarski A. A., Amati L., Mikolajewska J.,
  Belloni T., Del Sordo S., et al., 2001, ApJ, 561, 1006

	
\bibitem{} Gallo E., Fender R. P., Pooley G. G., 2003, MNRAS, 344, 60G

\bibitem{} Gallo E., Corbel S., Fender R. P., Maccarone T. J., Tzioumis A. K., 2004, MNRAS, 347, L52

\bibitem{} Gelino D. M., Balman S., Kiziloglu U., Yilmaz A., Kalemci  E., Tomsick J. A., 2006, ApJ, 642, 438G

\bibitem{} George I. M., Fabian A. C., 1999, MNRAS, 249, 352


\bibitem{} Gierlinski M., Zdziarski A. A.. Poutanen J., Coppi P. S., Ebisawa K., Johnson W. N., 1999, MNRAS, 309, 496G
	
\bibitem{} Gierlinski M., Done C., Page K., 2008, MNRAS, 388, 753G
	

\bibitem{} Greene, J., Bailyn, C. D., and Orosz, J. A., 2001, Apj,   554, 1290
		

\bibitem{} Hjellming R. M., Rupen M. P., 1995, Nature, 375, 464H

\bibitem{} Homan J., Wijnands R., Kong A., Miller J. M., Rossi S.,  Belloni T., Lewin W. H. G., 2006, MNRAS, 366, 235H

\bibitem{} Hynes R.I., Steeghs D., Casares J., Charles P.A., \& O'Brian K., 2003, ApJ, 583, L95

\bibitem{} Koyama K., et al. 2007, PASJ, 59S, 23K
	
\bibitem{} Laor, A., 1991, ApJ, 376, 90

	
\bibitem{} Maccarone T. J., 2003, A\&A, 409, 697M

\bibitem{} Makishima K., Takahashi H., Yamada S., Done C., Kubota A.,   Dotani T., Ebisawa K., Itoh T., Kitamoto S., Negoro H., Ueda Y.,   Yamaoka K., 2008, PASJ, 60, 585M

\bibitem{} Markoff S., Nowak M. A., 2004, ApJ, 609, 972

\bibitem{} Markoff S., Nowak M. A., \& Wilms J., 2005, ApJ, 635, 1203

\bibitem{} Martin R. G., Tout C.A., Pringle, J. E., 2008, MNRAS, 387,  188M

\bibitem{} Martin R. G., Reis R. C., Pringle, J. E., 2008, MNRAS,   391L, 15M

\bibitem{} Massey P., Johnson K. E., Degioia-Eastwood K., 1995, ApJ,   454, 151M
	

\bibitem{} McClintock J.E.,Haswell C. A., Garcia M. R., Drake J. J.,   Hynes R. I., Marshall H. L., Muno M. P., et al., 2001, ApJ, 555, 477

\bibitem{} McClintock J. E., Remillard R. A., 2006, in Lewin W. H. G., van der Klis M., eds, Compact Stellar X-ray Sources.Cambridge Univ. Press, Cambridge, p. 157


\bibitem{} Merloni A., Fabian, A. C., Ross, R. R., 2000, MNRAS, 313,   193 

\bibitem{} Merloni A., Di Matteo T., Fabian A. C., 2000, MNRAS, 318L, 15M 

\bibitem{} Merloni A., Fabian, A. C., 2001, MNRAS, 332, 165

\bibitem{} Merloni A., Di Matteo T., Fabian A. C., 2001, ApSSS, 276, 213M

\bibitem{} Merloni A., Fabian, A. C., 2002, MNRAS, 332, 165


\bibitem{} Miller J. M., Fabian A. C., Wijnands R., Reynolds C. S.,   Ehle M., Freyberg M. J., van der Klis M., Lewin W. H. G.,   Sanchez-Fernandez C., Castro-Tirado A. J., 2002, ApJ, 570L, 69M

\bibitem{} Miller J.M., Fabian A.C., Reynolds C.S., Nowak M.A., Homan   J., et al. 2004, ApJ, 606, L131

\bibitem{} Miller J. M., Fabian A. C., Nowak M. A., Lewin W. H. G.,   2005, tmgm.meet., 1296M

\bibitem{} Miller J.M., Homan J., Steeghs D., Rupen M., Hunstead R.W., Wijnands R., Charles P.A., \& Fabian A.C., 2006, ApJ, 653, 525

\bibitem{} Miller J. M., Homan J., and Miniutti G., 2006, ApJ, 652,   L113 

\bibitem{} Miller J. M., 2007, ARA\&A, 45, 441

\bibitem{} Miller J.M., Reynolds C.S., Fabian A.C., Cackett E.M., Miniutti G., Raymond J. Steeghs D., Reis R. C., Homan J., 2008, ApJ, 679L, 113M

	
\bibitem{} Miller J. M., Reynolds C. S., Fabian A. C., Miniutti G., Gallo L. C., 2009, ApJ, 697, 900M

\bibitem{} Miller J. M., 2009, ATel, 1966, 1M

\bibitem{} Miller J. M., Cackett E. M.,\&  Reis R. C., 2009, ApJ accepted  arXiv0910.2877M

\bibitem{} Miniutti G., Fabian A. C., Miller J. M., 2004, MNRAS, 351,   466


\bibitem{} Mitsuda K., Inoue H., Koyama K., Makishima K. Matsuoka M.,  Ogawara Y., Suzuki K.,et al., 1984, PASJ, 36,   741

\bibitem{} Munos-Darias T., Casares J., Martinez-Pais I.G., 2008,  MNRAS, arXiv:0801.3268v1 [astro-ph]


\bibitem{} Murdin P., Webster L. B., 1971, Nature 233, 110 

\bibitem{} Narayan R., \& Yi I., 1995, ApJ, 452, 710

\bibitem{} Narayan R., Yi I., \& Mahadevan, R., 1995, Nature, 374, 623N

\bibitem{} Narayan R., Mahadevan R. Grindlay J. E., Popham R. G., Gammie C., 1998, ApJ, 492, 554N

\bibitem{} Orosz J. A., McClintock J. E., Remillard R. A., Corbel S.,   2004, ApJ, 616, 376O

	
\bibitem{} Paizis A., Nowak M. A., Chaty S., Rodriguez J.,   Courvoisier T. J.-L., Del Santo M., Ebisawa K., Farinelli R.,   Ubertini P., Wilms J., 2007, ApJ, 657L, 109P

	
\bibitem{} Paizis A., Ebisawa K., Takahashi H., Dotani T., Kohmura T.,   Kokubun M., Rodriguez J., Ueda Y., Walter R., Yamada S., Yamaoka,   K., Yuasa T., 2008, arXiv0811.2663P

\bibitem{} Ramadevi, M. C., and Seetha, S., 2007, MNRAS, 378, 182 


\bibitem{} Reis R. C., Fabian A. C., Ross R. R., Miniutti G.,  Miller J. M., Reynolds C., 2008, MNRAS, 387, 1489R

\bibitem{} Reis R. C., Miller J. M., Fabian A. C., 2009a, MNRAS, 395L,   52R

\bibitem{} Reis R. C., Fabian A. C., Ross R. R., Miller J. M., 2009b,   MNRAS, 395, 1257R

\bibitem{} Reis R. C., Fabian A. C., Young A. J., 2009c, MNRAS in press (arXiv0904.2747R)
	
\bibitem{} Remillard R., Levine A. M., Morgan E. H., Markwardt C. B.,   Swank J. H., 2006, ATel, 714, 1R

\bibitem{} Romero G. E., Kaufman Bernado M. M., Mirabel I. F., 2002,   A\&A, 393L, 61R

\bibitem{} Ross R. R., Fabian A. C., \& Mineshige S., 1992, MNRAS, 258, 189

\bibitem{} Ross R.R., \& Fabian A. C., 1993, MNRAS, 261, 74

\bibitem{} Ross R.R., Fabian A. C., \& Young A. J., 1999, MNRAS 306, 461R

\bibitem{} Ross R.R., \& Fabian A. C., 2005, MNRAS, 358, 211


\bibitem{} Rossi S., Homan J., Miller J. M., Belloni T., 2005, MNRAS, {\thinspace 360, 763}

\bibitem{} Rykoff E. S., Miller J. M.,Steeghs D.,Torres M. A. P.,   2007, ApJ, 666, 1129R (R07)

\bibitem{} Sala G., Greiner J., Ajello M., Bottacini E., Haberl  F., 2007, A\&A, 473, 561S


\bibitem{} Shakura N.~I. and Sunyaev R.~A., 1973, A\&A, 24, 337

\bibitem{} Shapiro S. L., Lightman A. P., Eardley D. M., 1976, ApJ, 204, 187S

\bibitem{} Shaposhnikov N., Titarchuk L., 2009, ApJ, 699, 453S

\bibitem{} Shimura, T., Takahara F., 1995, ApJ, 445, 780S

\bibitem{} Stirling A. M., Spencer R. E., de la Force C. J.,   Garrett M. A., Fender R. P., Ogley, R. N., 2001, MNRAS, 327, 1273S

\bibitem{} Sunyaev R. A., Titarchuk L. G., 1980, A\&A, 86, 121S

\bibitem{} Takahashi K., Inoue H. Dotani T., 2001, PASJ, 53, 1171T

\bibitem{} Takahashi H., et al. 2008, PASJ, 60S, 69T

\bibitem{} Tanaka Y., Nandra K., Fabian A. C., Inoue H., Otani C., Dotani T., Hayashida K., Iwasawa K., Kii T., Kunieda H., Makino F., Matsuoka M., 1995, Nature, 375, 659T

\bibitem{} Tananbaum H., Gursky H., Kellogg E., Giacconi R., Jones C.,   1972, ApJ, 177L, 5T

\bibitem{} Thorne K. S., 1974, ApJ, 191, 507

\bibitem{} Torres D. F., Romero G. E., Barcons X., Lu Y., 2005, ApJ,   626, 1015T

\bibitem{}Wagner R. M., Foltz C. B., Shahbaz T., Casares J., Charles   P. A., Starfield S. G., Hewett P., 2001, ApJ, 556, 42
	
\bibitem{} Wilms J., Nowak M. A., Dove J. B., Fender R. P., Di Matteo   T., 1999, ApJ, 522, 460

\bibitem{} Young A. J., Fabian A. C., Ross R. R., Tanaka Y., 2001,   MNRAS, 325, 1045Y

\bibitem{} Zdziarski A.A., Gierlinski M., Mikolanjewska J., Wardzinski
  G., Smith D.M., Harmon A., \&\ Kitamoto S., 2004, MNRAS, 351, 791

\bibitem{} Zdziarski A.A., Gierlinski M.,2004, PThPS, 155, 99Z

\bibitem{} Zimmerman E. R., Narayan R., McClintock J. E., Miller
  J. M., 2005, ApJ, 618, 832Z

\bibitem{} Ziolkowski J., 2005, MNRAS, 358, 851Z
	

\bibitem{} Zurita C., Durant M., Torres M. A. P., Shahbaz T.,
  Casares J., Steeghs D., 2008, ApJ, 681, 1458Z

\end{thebibliography}
\end{document}